\documentclass[11pt,a4paper]{article}
\pdfoutput=1
\usepackage{jheppub}
\usepackage{graphicx,amssymb,amsmath,amsfonts}
\usepackage[english]{babel}
\usepackage{slashed}
\usepackage{multirow}

\usepackage{braket}
\usepackage{simplewick}
\usepackage{pdfpages}
\usepackage{caption}
\usepackage{subcaption}
\usepackage{cleveref}
\usepackage{enumitem}

\newcommand{\mn}{{\mu\nu}}
\newcommand{\rs}{{\rho\sigma}}
\newcommand{\mnrs}{{\mu\nu\rho\sigma}}
\newcommand{\rsmn}{{\rho\sigma\mu\nu}}

\newcommand{\has}{\mathcal{H}_\text{as}}
\newcommand{\hf}{\mathcal{H}_\text{F}}

\newcommand{\V}{\mathbf}

\newcommand{\wk}{\omega_{k}}

\newcommand{\tdk}[1]{{\frac{d^3{#1}}{(2\pi)^3 2\omega_{#1}}}}
\newcommand{\tdp}[1]{{\frac{d^3{#1}}{(2\pi)^3 2\omega_{#1}}}}
\newcommand{\td}[1]{\widetilde{d^3 #1}\,}
\newcommand{\Afactor}{A}
\newcommand{\Qfin}{\widetilde{\mM}}
\newcommand{\Qfintilde}{\widetilde{\mM}'}

\def\mL{\mathcal{L}}
\def\mS{\mathcal{S}}

\def\mP{\mathcal{P}}

\def\mM{\mathcal{M}}

\def\mD{\mathcal{D}}
\def\mM{\mathcal{M}}
\def\mS{\mathcal{S}}

\def\mI{\mathcal{I}}

\def\zb{\bar{z}}

\def\omegab{\bar{\omega}}

\def\pa{\partial}

\def\g5{\gamma_5}

\def\lamt{\tilde{\lambda}}

\def\b[#1]{\bold{#1}}
\def\bb[#1]{\overline{\bold{#1}}}

\def\bs[#1,#2]{\bold{#1}_{#2}}
\def\bbs[#1,#2]{\overline{\bold{#1}}_{#2}}

\def\s2{\sigma_2}

\def\ep{\epsilon}

\def\gammaflat{ \gamma_{z\zb}}
\def\gammaflatk{ \gamma_{z_k \zb _k}}
\def\Tsoft{T_{\text{soft}}}
\def\Rfp{R_f({p})}
\def\omegak{\omega_k}
\def\omegap{\omega_p}
\def\paz{\pa_z}
\def\pazU{\pa^z}
\def\pazb{\pa_{\zb}}
\def\pazbU{\pa^{\zb}}
\def\Mksoft{\mM_{k, \,\text{soft}}}

\makeatother

\title{Asymptotic Dynamics in Perturbative Quantum Gravity and BMS Supertranslations}
\author{Sangmin Choi, Uri Kol and Ratindranath Akhoury}

\affiliation{Michigan Center for Theoretical Physics, \\
Randall Laboratory of Physics, Department of Physics,\\
University of Michigan, Ann Arbor, MI 48109, USA}

\emailAdd{sangminc,urikol,akhoury@umich.edu}

\abstract{

Recently it has been shown that infrared divergences in the conventional S-matrix elements of gauge and gravitational theories arise from a violation of the conservation laws associated with large gauge symmetries.
These infrared divergences can be cured by using the Faddeev-Kulish (FK) asymptotic states as the basis for S-matrix elements.
Motivated by this connection, we study the action of BMS supertranslations on the FK asymptotic states of perturbative quantum gravity.
We compute the BMS charge of the FK states and show that it characterizes the superselection sector to which the state belongs.
Conservation of the BMS charge then implies that there is no transition between different superselection sectors, hence showing that the FK graviton clouds implement the necessary vacuum transition induced by the scattering process.

}

\begin{document}
\maketitle

\section{Introduction}
Since the pioneering work of Bloch and Nordsieck \cite{BN} it is known that in quantum electrodynamics, a charged particle may emit an infinite number of soft photons with a finite total energy below the energy resolution of the detectors. This gives rise, order by order in perturbation theory, to infrared divergences in real emission processes which are known to exponentiate \cite{Weinberg:1965nx}. The virtual diagrams are also infrared divergent and these contributions exponentiate as well, implying the vanishing of the corresponding S-matrix elements as the infrared cut-off is removed. The standard way to deal with these divergences is to sum up physically indistinguishable (inclusive) cross-sections, order by order in perturbation theory, which leads to their cancellation - the infrared cut-off being replaced by the detector resolution. This Bloch-Nordsieck mechanism for the cancellation of infrared divergences works also in perturbative quantum gravity and for appropriate processes in non-abelian gauge theories. Even though physically sensible results are obtained in this way, a troubling feature is that the S-matrix in the standard Fock basis is not defined. 

In \cite{Kulish:1970ut}, Faddeev and Kulish developed a formalism which allows for the construction of appropriately defined S-matrix elements. This takes into account the fact that the early and late time dynamics in theories with massless gauge particles cannot be free. For example, in QED, they observed that there are terms in the interaction Hamiltonian that arise from the coupling of soft photons to creation or annihilation operators of a charged particle and give a non-vanishing contribution in the limit $t\rightarrow \pm\infty$. The proposed solution was to construct true asymptotic states which include multiple soft particle emission to all orders in the coupling constant. Physically, these states describe dressing of the charged particles by soft photon clouds. The standard Dyson S-matrix between these asymptotic states is then free of infrared singularities. This Faddeev-Kulish method was extended to perturbative quantum gravity in \cite{Ware:2013zja}. It should be noted that the Faddeev-Kulish construction is valid in QED only for massive charged particles. In contrast, perturbative quantum gravity does not suffer from this restriction because of the cancellation of collinear divergences \cite{Akhoury:2011kq}. In this sense, the infrared behavior of gravity is simpler than that of QED. 

Studies of the large time structure of gauge and gravitational theories have recently seen a resurgence following the original work of \cite{Bondi:1962px,Sachs:1962wk,Christodoulou:1993uv}. In particular, an important realization has emerged that the infrared sector of these theories is governed by an infinite dimensional symmetry group generated by large gauge transformations \cite{Strominger:2013jfa,He:2014laa,He:2014cra,Kapec:2015ena}. In perturbative gravity, for example, there are infinitely degenerate vacua which differ by the addition of soft gravitons and are related by spontaneously broken Bondi-van der Burg-Metzner-Sachs (BMS) symmetries. The infinite number of conservation laws associated with the BMS supertranslations forbids the transitions between equivalent vacua, and this is interpreted (for the case of QED see \cite{Gabai:2016kuf,Kapec:2017tkm}) as the reason of the vanishing of the Fock space S-matrix elements for transitions involving soft gravitons. We have seen that the Faddeev-Kulish (FK) asymptotic states were introduced to precisely take care of this problem. Since the Dyson S-matrix is finite between the FK asymptotic states, the question that naturally arises is, what is the relation between the FK asymptotic operator which generates the asymptotic states and the BMS supertranslations? For the aforementioned interpretation to be valid, it must be that the FK dressings implicitly induce transitions between the degenerate vacua. This interpretation has been explicitly verified for the case of QED in \cite{Gabai:2016kuf} and \cite{Kapec:2017tkm}. In this paper we extend and generalize their analysis to the case of perturbative quantum gravity.

Related discussions have appeared in \cite{Mirbabayi:2016axw,Bousso:2017dny,Bousso:2017rsx}, where the authors study the factorization of the hard and soft sectors in a scattering experiment, and the relation to BMS transformations, with a special emphasis on the application of their results to the black hole information paradox.
In this paper, however, we study only the soft dynamics in the asymptotic region of Minkowski spacetime.
We would also like to mention the work of \cite{Carney:2017jut}, in which the relation between the soft and hard sectors was formulated using information-theoretic tools.

The paper is organized as follows. In section \ref{AsymStates} we review and generalize the FK construction for perturbative quantum gravity. In particular, the construction of the asymptotic states for arbitrary values of the BMS charge is outlined, and the cancellation of infrared divergences for the Dyson S-matrix between these asymptotic states is explicitly shown for arbitrary number of external matter particles. A requirement for this cancellation is that the BMS charge for incoming and outgoing states must be the same. Certain technical details are relegated to the appendices. In section \ref{BMS}, we review the structure of BMS transformations and the construction of the generator of supertranslations. In section \ref{ActionOfBMS}, we consider the action of BMS supertranslations on the vacuum state and on undressed matter particles. Using these results, in section \ref{DressedParticles} we consider BMS supertranslations acting on a FK asymptotic state. We show that the BMS supertranslations acting on a bare particle and on the momentum-dependent part of its accompanying graviton cloud exactly cancel each other, thus verifying the consistency of the relation between the FK states and the action of the BMS supertranslations referred to in the previous paragraph.

\section{Physical asymptotic states for gravity}\label{AsymStates}

Kulish and Faddeev have constructed physical asymptotic states in QED \cite{Kulish:1970ut}, following the works of Chung \cite{Chung:1965zza} and Kibble \cite{Kibble:1969ip}. The physical asymptotic states were constructed by dressing the incoming and outgoing states with a coherent cloud of photons.
This formalism has been used in \cite{Ware:2013zja} to construct asymptotic states in gravity, which involve coherent clouds of gravitons.
Since we will be working with these states extensively, in this section we present a brief overview of the work done in \cite{Ware:2013zja}
	and provide generalizations that are relevant to our discussion.

Throughout this paper we will be working in the leading soft approximation, the eikonal limit, when the spin of the matter particles does not play a role. Thus, our results are equally valid for scalar or fermionic matter; however, to be specific, we will explicitly work with a massive scalar field $\varphi$ coupled to gravity.
By expanding the metric about flat space,
\begin{align}
g_\mn=\eta_\mn+\kappa h_\mn,
\end{align}
where $\eta_\mn=\text{diag}(-1,1,1,1)$ and $\kappa^2=32\pi G$, one obtains the leading-order interaction Lagrangian
\begin{align}
\label{eqn:Lint}
\mL_\text{int}=\frac{\kappa}{2}
\left[
	h^\mn\partial_\mu\varphi\partial_\nu\varphi
	- \frac{h}{2}
	\left(
		\partial^\mu\varphi\partial_\mu\varphi + m^2\varphi^2
	\right)
\right],
\end{align}
with $h=h_\mu^\mu=\eta^\mn h_\mn$.
The fields $\varphi(x)$ and $h_\mn(x)$ can be expanded in harmonic modes
\begin{gather}
\label{eqn:phi}
\varphi(x)=\int \td{p}
	\left(
		b(p) e^{ip\cdot x}
		+ b^\dagger(p) e^{-ip\cdot x}
	\right)\\
\label{eqn:hmn}
h_\mn(x)= \int \td{k}
	\left(
		a_\mn(k) e^{ik\cdot x}
		+ a^\dagger_\mn(k) e^{-ik\cdot x}
	\right),
\end{gather}
where we have employed the shorthand notation
\begin{align}
\label{tilde_shorthand}
\td{p}=\tdp{p},\quad\td{k}=\tdk{k},
\end{align}
with $\omegap=\sqrt{|\V{p}|^2+m^2}$ and $\wk=|\V{k}|$.
With our choice of normalization (which is different from \cite{Ware:2013zja}), the creation and annihilation operators obey the commutation relations
\begin{align}
\label{eqn:bcomm}
\left[b(p),b^\dagger(p')\right]
	&= (2\pi)^3 (2\omegap) \delta^3(\V{p}-\V{p'})
\\
\label{eqn:acomm}
\left[a_\mn(k),a_\rs^\dagger(k')\right]
	&= \frac{1}{2}I_\mnrs(2\pi)^3 (2\wk) \delta^3(\V{k}-\V{k'}),
\end{align}
where
\begin{align}
	I^\mnrs \equiv \eta^{\mu\rho}\eta^{\nu\sigma} + \eta^{\mu\sigma} \eta^{\nu\rho} - \eta^\mn \eta^\rs.
\end{align}
We choose to work in the harmonic gauge (also known as the de Donder gauge)
\begin{align}
	\label{HarmonicGauge}
	\partial^\mu h_\mn - \frac{1}{2}\partial_\nu h = 0.
\end{align}
We define the physical states as the subset of all the states in the Fock space that obey the gauge condition \eqref{HarmonicGauge}.
The gauge condition \eqref{HarmonicGauge} translates into the Gupta-Bleuler constraint on the Fock space
\begin{align}
\label{Gupta_Fock}
\left(
	k^\mu a_\mn(k) - \frac{1}{2}k_\nu a^\mu_\mu(k)
\right)
\ket{\Psi}=0
\quad\text{for all physical Fock states $\ket{\Psi}$}.
\end{align}
In the following we will discuss the construction and properties of physical asymptotic states.

In \cite{Ware:2013zja}, the Faddeev-Kulish formalism was used to construct an operator $e^{R(t)}$ that projects the Fock space $\hf$ into the space of asymptotic states $\has$
\begin{align}
	e^{R(t)}\hf = \has.
\end{align}
The anti-Hermitian operator $R(t)$ is given by
\begin{align}
R(t)=\frac{\kappa}{2}
	\int\td{p}\td{k} \rho(p)
	\frac{p^\mu p^\nu}{p \cdot k}
	\left(
		a^\dagger_\mn(k) e^{-i\frac{p\cdot k}{\omegap}t}
		- a_\mn(k) e^{i\frac{p\cdot k}{\omegap}t}
	\right),
\end{align}
where $t$, the asymptotic time, is taken to be very large and
$\rho(p)=b^\dagger(p)b(p)$ is the number operator of the scalar particle.
For QED \cite{Kulish:1970ut} and for gravity \cite{Ware:2013zja}, it was shown that this asymptotic space has a number of important properties including gauge invariance (linearized general coordinate invariance). However, recent works \cite{Strominger:2013jfa,He:2014laa} have clarified that the gauge invariance is only with respect to small gauge transformations. Large gauge transformations, or those that do not reduce to the identity at time-like and null infinity, are instead symmetries of the system. Among these, the ones relevant for us are the BMS supertranslations. The space of asymptotic states is divided into superselection sectors, each labelled by a BMS charge which is explicitly constructed in section \ref{DressedParticles}. One problem that arises is that in each superselection sector the operator $e^{R(t)}$ acting on the Fock states creates an unbounded number of low energy gravitons. Noting that only the low energy behavior of $R(t)$ defines the space $\has$, we may introduce another operator $R_f$ of the form:
\begin{align}
R_f=\frac{\kappa}{2}
	\int\td{p}\td{k} \rho(p)
	\left(
		f^{\mn*}(p,k) a^\dagger_\mn(k)
		- f^\mn(p,k) a_\mn(k)
	\right),
\end{align}
which is characterized by an infrared function $f_\mn(p,k)$. This function will be different in different superselection sectors, but its form is restricted as we will show now. The restrictions come from the fact the $e^{R(t)}$ and $e^{R_f}$ must describe unitarily equivalent spaces, i.e.,
\begin{align}
e^{R_f}\hf =e^{R(t)}\hf \nonumber = \has.
\end{align}
The constraints arising from these, which are discussed in \cite{Ware:2013zja} and in \cref{app_c}, make it convenient to write $f_\mn(p,k)$ in the form
\begin{align}
\label{eqn:fmn}
f_\mn(p,k)=\left[\frac{p_\mu p_\nu}{p\cdot k} + \frac{c_\mn(p,k)}{\wk}\right]\phi(p,k)
\end{align}
for some function $c_\mn(p,k)$, where $\phi(p,k)$ is a smooth function such that $\phi\rightarrow 1$ as $k\rightarrow 0$.
Depending on how we parametrize $c_\mn(p,k)$, \eqref{eqn:fmn} might not be compatible with some
	possible forms of $f_\mn(p,k)$; for example the last parametrization used in \cite{Ware:2013zja}, which reads
\begin{align}
\label{eqn:f_old_param}
f_\mn(p,k)
	=
	\frac{p^\rho p^\sigma}{p\cdot k}\epsilon^-_\rs(k)\epsilon^+_\mn(k)
	+ \frac{p^\rho p^\sigma}{p\cdot k}\epsilon^+_\rs(k)\epsilon^-_\mn(k),
\end{align}
where $\epsilon^\pm_\mn(k)$ are the transverse, traceless physical polarization tensors of graviton:
\begin{equation}\label{TTcomponents}
k^\mu\ep^{\pm}_\mn(k) = 0
	\quad\text{and}\quad
	\eta^\mn\ep^{\pm}_\mn(k)=0.
\end{equation}
However, since we will use explicit parameterization only as an example, our results will be valid in general.

In addition, physical asymptotic states are subject to the Gupta-Bleuler condition \eqref{Gupta_Fock}, which implies
\begin{align}
\left[R_f,k^\mu a_\mn - \frac{1}{2}k_\nu a^\mu_\mu\right]=0,
\end{align}
or,
\begin{align}
\label{eqn:c1}
k^\mu f_\mn =
\frac{k^{\mu} c_{\mu\nu}}{\omega_k} +p_{\nu}
= 0.
\end{align}
There are additional constraints on the function $f_\mn$, or equivalently, on $c_\mn$ arising again from the fact that $e^{R(t)}$ and $e^{R_f}$ define unitarily equivalent spaces. In \cref{app_c} we show that, to leading order in $k$, these constraints are
\begin{gather}
\label{eqn:cc2}
c^*_\mn(p,k)=c_\mn(p,k)\\
\label{eqn:cc3}
c_\mn(p,k) I^\mnrs c_\rs(p',k)=0 \quad\text{for all $p$ and $p'$}.
\end{gather}
Subleading corrections to \eqref{eqn:cc2}-\eqref{eqn:cc3} will only rescale the operator $e^{R_f}$ by a positive finite constant, and we could therefore absorb them in the normalization of the state.

With a $c_\mn$ that satisfies \eqref{eqn:c1}-\eqref{eqn:cc3}, the graviton cloud operator $e^{R_f}$
	properly gives us the asymptotic states
\begin{align}
\label{asymptotic-space}
\ket{\Psi^\text{as}}=e^{R_f}\ket{\Psi},
\end{align}
where $\ket{\Psi}$ denotes Fock states.
It is convenient to interpret $e^{R_f}$ as an operator that dresses each scalar with its own cloud of gravitons. Indeed, this is seen most clearly by
commuting $e^{R_f}$ through the scalar operators
	using $\left[ b(p), \rho(p') \right]=(2\pi)^3 (2 \omegap) \delta^3 (\V{p}- {\V{p'}})b(p)$. In this way we obtain, for example
\begin{equation}
e^{R_f}  b^\dagger(p_1) b^\dagger(p_2) \ket{0} = e^{R_f(p_1)} b^\dagger(p_1) e^{R_f(p_2)} b^\dagger(p_2) \ket{0},
\end{equation}
where
\begin{equation}
\Rfp= 
\frac{\kappa}{2}
\int \td{k}
\left(
f^{\mu\nu}(p,k) a_{\mu\nu}^{\dagger}(k)
-f^{\mu\nu}(p,k) a_{\mu\nu}(k)
\right).
\end{equation}

One can parameterize the c-matrix as the following to exclude terms proportional to $p_\mu p_\nu$,
\begin{equation}
\label{eqn:c_param}
c_{\mu\nu}(p,k) =
a_1 q_{(\mu}p_{\nu)}
+a_2 q_{\mu}q_{\nu}\ ,
\end{equation}
where $q(k)$ is some four-vector and $a_1$, $a_2$ are coefficients to be determined\footnote{
We have used the following notation for the symmetric combination $q_{(\mu}p_{\nu)} \equiv q_{\mu}p_{\nu}+q_{\mu}p_{\nu}$.}.
This parameterization is similar to the one used in \cite{Gabai:2016kuf} for the case of QED.
The gauge constraint \eqref{eqn:c1} then fixes the coefficients to be
\begin{equation}
\begin{aligned}
a_1 = - \frac{\wk}{k \cdot q}
\quad\text{and}\quad
a_2 = -\frac{k \cdot p}{k \cdot q}a_1
\ ,
\end{aligned}
\end{equation}
and therefore we have
\begin{equation}\label{cmatrix}
c_{\mu\nu}(p,k) =\frac{\omega_k}{k \cdot q}
\left[
\frac{k \cdot p}{k \cdot q}
q_{\mu}q_{\nu}
-q_{(\mu}p_{\nu)}
\right].
\end{equation}
The constraint \eqref{eqn:cc3} then reads
\begin{equation} \label{eqn:q_null}
c_\mn(p,k)I^\mnrs c_\rs(p',k)
=
	\frac{\wk^2}{(k\cdot q)^2}
		\,q^2
		\left[
			\frac{(k\cdot p)}{(k\cdot q)}q - p
		\right]
		\cdot
		\left[
			\frac{(k\cdot p')}{(k\cdot q)}q - p'
		\right]=0,
\end{equation}
and can be satisfied identically only if $q$ is a null vector $q^2=0$.
In addition, since rescaling $q$ by a constant does not affect \eqref{cmatrix}, we can assume that the time component of $q$ is $1$ without any loss of generality.
As we will see later, the null vector $q$ parameterizes the space of superselection sectors, and the combination
\begin{align}
	c^\mn(p,k) \ep^\pm_\mn(k)
\end{align}
is related to the conserved charge under BMS symmetry transformations. The BMS charge therefore characterizes the superselection sector.
A similar conclusion was drawn for QED in \cite{Gabai:2016kuf}.
In \cite{Ware:2013zja}, the choice $c^\mn(p,k) \epsilon^\pm_\mn(k) = 0$ was made. This choice can be realized by \eqref{cmatrix} with $q^\mu=(1,-\hat{\V{k}})$ and corresponds to a vanishing BMS charge.

The harmonic gauge condition \eqref{HarmonicGauge} does not fix the gauge completely. BMS transformations parameterize the residual leftover gauge freedom \cite{Bondi:1962px,Sachs:1962wk,Strominger:2013jfa,He:2014laa}, which is given by
\begin{equation}
\label{residual}
h_{\mu\nu}   \ \rightarrow\  h_{\mu\nu} + \pa_{\mu}\lambda _{\nu} + \pa_{\nu}\lambda _{\mu} ,
\end{equation}
with the gauge parameter $\lambda_{\mu}$ satisfying the Laplace equation
\begin{equation}
\label{LaplaceEq}
\square \lambda_{\mu}= 0,
\end{equation}
We review this in section \ref{BMS}. Here it is worth noting that under this residual gauge freedom the infrared function $f_{\mu\nu}$ transforms as
\begin{align}
\label{eqn:fmngauge}
f_\mn(p,k) \ \rightarrow\  f_\mn(p,k) + k_{(\mu}\lambda_{\nu)} - (k\cdot \lambda)\eta_\mn,
\end{align}
with $k^2=0$, which is also implied by equation \eqref{eqn:c1}. It was shown in \cite{Ware:2013zja} that the action of $e^{R_f}$ on a matter Fock state $\ket{\Psi_\text{in}}$ is invariant under \eqref{eqn:fmngauge} for small gauge transformations.
In the subsequent sections, we will scrutinize how the BMS transformation, in particular the supertranslation,
	plays a role in the context of the asymptotic states.

We would now like to show that the S-matrix elements in the basis \eqref{asymptotic-space}
\begin{align}
\bra{\Psi^\text{as}_\text{out}}\mS\ket{\Psi^\text{as}_\text{in}}
=
\bra{\Psi_\text{out}}e^{-R_f}\mS e^{R_f}\ket{\Psi_\text{in}},
\end{align}
are free of IR divergence.
This has been shown in \cite{Ware:2013zja} for a process between
	single-scalar asymptotic states using a specific choice of $c_\mn$.
In \cref{app_ir}, we present a generalization of this result;
we show that the divergences cancel for a process between asymptotic states with arbitrary number of
	scalar particles to all orders in the perturbative expansion,
	provided that the $c$-matrices satisfy, to leading order in the momentum $k$,
\begin{align}
\label{ctot_must_die_brief}
\sum_{j\in\text{out}}c^\text{(out)}_\mn(p_j,k)=\sum_{i\in\text{in}}c^\text{(in)}_\mn(p_i,k),
\end{align}
where ``in" and ``out" denote the set of incoming and outgoing scalar particles, respectively.
We briefly describe these results now.

First, note that from the work of Weinberg \cite{Weinberg:1965nx} we know that the amplitude $\mM$ of a process can be decomposed into
\begin{align}
\mM = \bra{\Psi_\text{out}}e^{-R_f}\mS e^{R_f}\ket{\Psi_\text{in}}=\Afactor_\text{virt}\mM',
\end{align}
where $\Afactor_\text{virt}$ is the IR-divergent contribution of virtual gravitons and $\mM'$ is the remainder of the amplitude.
In \eqref{eqn:acloud_factor} we show that
\begin{align}
\mM'=\Afactor_\text{cloud}\Qfin,
\end{align}
where $\Qfin$ is the IR-finite part of the amplitude
	and $\Afactor_\text{cloud}$ is the divergent factor coming from interactions that involve graviton clouds.
	The latter has the form
\begin{align}
\Afactor_\text{cloud}=(\Afactor_\text{virt})^{-1} e^{-a C},
\end{align}
where $a$ is a positive constant, and
\begin{align}
\label{eqn:eac_ctot}
C\equiv \int\frac{d^3k}{\wk^3}\,
	c^\text{tot}_\mn
	I^\mnrs
	c^\text{tot}_\rs
\quad\text{with}\quad
	c^\text{tot}_\mn=\sum_{j\in\text{out}}c_\mn(p_j,k)-\sum_{i\in\text{in}}c_\mn(p_i,k).
\end{align}
The factor $e^{-aC}$ derives solely from the interactions between graviton clouds.
Since we have the same $c$-matrices for both the incoming and outgoing states, by \eqref{eqn:cc3}
	the integrand of \eqref{eqn:eac_ctot} vanishes and $C=0$.
If we use different $c$-matrices, for instance $c_\mn$ for incoming and
	$c'_\mn$ for outgoing states, then \eqref{eqn:eac_ctot} readily generalizes to the same expression for the integral with
\begin{align}
	c^\text{tot}_\mn=\sum_{j\in\text{out}}c'_\mn(p_j,k)-\sum_{i\in\text{in}}c_\mn(p_i,k).
\end{align}
If the condition \eqref{ctot_must_die_brief} is not met, then $C$ exhibits IR divergence and the amplitude will vanish.
Therefore to obtain a non-zero amplitude, \eqref{ctot_must_die_brief} must be satisfied and $C=0$.
It is worth noting that subleading corrections in the momentum $k$ to equation \eqref{ctot_must_die_brief} are finite and can therefore be absorbed in the normalization of the states.
The amplitude thus becomes
\begin{align}
\mM = \Afactor_\text{virt}\Afactor_\text{cloud}\Qfin
	= \Afactor_\text{virt}(\Afactor_\text{virt})^{-1}e^{-\alpha C}\Qfin
	= \Qfin,
\end{align}
which is IR finite.

\section{The BMS group}\label{BMS}

\subsection{Asymptotically flat spacetime}

In this section we review the structure of asymptotically Minkowski geometry and BMS transformations.
We will follow closely the works of \cite{He:2014laa,Strominger:2013jfa}.

Let us first define the retarded system of coordinates, which is related to the Cartesian system by
\begin{equation}
r^2 = x_1^2 +x_2^2 +x_3^2, \qquad 
u=t-r, \qquad
z = \frac{x_1+ i x_2}{r+x_3}.
\end{equation}
The inverse relations are given by
\begin{equation}
t= u+ r\ , \qquad 
{\bold x} = r \hat{{\bold x}} = \frac{r}{1+z \zb} \left(z+\zb,i(\zb-z) , 1-z \zb   \right).
\end{equation}
The flat Minkowski metric is then given by
\begin{equation}
\begin{aligned}
ds^2_0  &=  -dt^2 +dx_1^2 +dx_2^2 +dx_3^2 \\
& =-du^2 -2 du dr +2r^2 \gamma _{z\zb} dz d\zb  
\ ,
\end{aligned}
\end{equation}
where
\begin{equation}
\gamma_{z\zb}=\frac{2}{\left( 1+z \zb \right)^2}
\end{equation}
is the round metric on the unit $S^2$.

Asymptotically flat metrics have an expansion around future null infinity ($r=\infty$), whose leading order terms
	are given by
\begin{equation}\label{asymMetric}
\begin{aligned}
ds^2  &= ds^2_0  \\
&+
\frac{2m_B}{r} du^2
+r C_{zz}dz^2
+r C_{\zb\zb}d\zb^2
-2U_z du dz 
-2 U_{\zb}du d\zb\\
&+\ \dots\ ,
\end{aligned}
\end{equation}
where
\begin{equation}
U_z = -\frac{1}{2} D^z C_{zz}\ ,
\qquad
U_{\zb} = -\frac{1}{2} D^{\zb} C_{\zb \zb}\ ,
\end{equation}
and the dots denote higher order terms.
The Bondi mass aspect $m_B$ and the radiative data $C_{zz}$, $C_{\zb\zb}$ are functions of $(u,z,\zb)$.
We also define the Bondi news by
\begin{equation}\label{Nzz}
N_{zz} \equiv \pa_u C_{zz}\ ,
\qquad
N_{\zb\zb} \equiv \pa_u C_{\zb\zb}
\ .
\end{equation}

The $\mI^{+}$ data $m_B$ and $C_{zz}$ are related by the constraint equation
\begin{equation}\label{constraintEq}
\pa_u m_B = -\frac{1}{2} \pa_u \left[ D^z U_z +D^{\zb} U_{\zb} \right] - T_{uu}
\ ,
\end{equation}
where
\begin{equation}
\label{eqn:T_uu}
 T_{uu} = \frac{1}{4} N_{zz}N^{zz} + 4\pi G \lim_{r \rightarrow \infty} \left[ r^2 T^M_{uu} \right]
\end{equation}
is the total outgoing radiation energy flux.
The first term of \eqref{eqn:T_uu} is the gravitational contribution while $T^M$ is the stress-energy tensor of the matter sector.

It is important to note that the metric in \eqref{asymMetric} is written in the Bondi gauge,
	which is convenient for the presentation of the asymptotic solution
	but is not compatible with the harmonic gauge.
The transformation that relates the two gauges
\begin{equation}
h_{\mu\nu}^{H}  = h_{\mu\nu}^{B} +\pa_{\mu}\xi_{\nu}+\pa_{\nu}\xi_{\mu}
\end{equation}
obeys the following equation
\begin{equation}\label{xiEq}
\square \xi _{\mu}
=
\frac{1}{2} \pa_{\mu} h^B - \pa^{\nu} h^B_{\mu \nu}
\end{equation}
where the label H stands for Harmonic gauge and the label B stands for Bondi gauge.
For a detailed discussion on the relation between the two gauges we refer the reader to references
\cite{Campiglia:2015kxa,Avery:2015gxa,Avery:2015rga}.
In the rest of the paper we will be working the harmonic gauge.

\subsection{BMS supertranslations}\label{GaugeModes}

As discussed in the previous section, after fixing the harmonic gauge there is still a residual leftover gauge freedom given by \eqref{residual}-\eqref{LaplaceEq}.
The gauge field $\lambda_{\mu}$ parameterizes the group of BMS transformations. At leading order it is given by \cite{Campiglia:2015kxa}
\begin{equation}
\lambda^{\mu} \pa_{\mu} = f\pa_{u}
+ V^i \pa_i  + \frac{1}{2} (D^i V_i ) 
\left( u \pa_u 
- r  \pa_r
\right)
+\dots
\ ,
\end{equation}
where $i=1,2$ runs over the $S^2$ coordinates, and the dots stand for subleading terms.
The function $f(z,\zb)$ is the transformation parameter of supertranslations, and the 2-vector $V^i(z,\zb)$ is the transformation parameter of super-rotations.
In this paper we will be interested only in the supertranslations.

\begin{figure}[t]
\centering
\includegraphics[width=.49\textwidth]{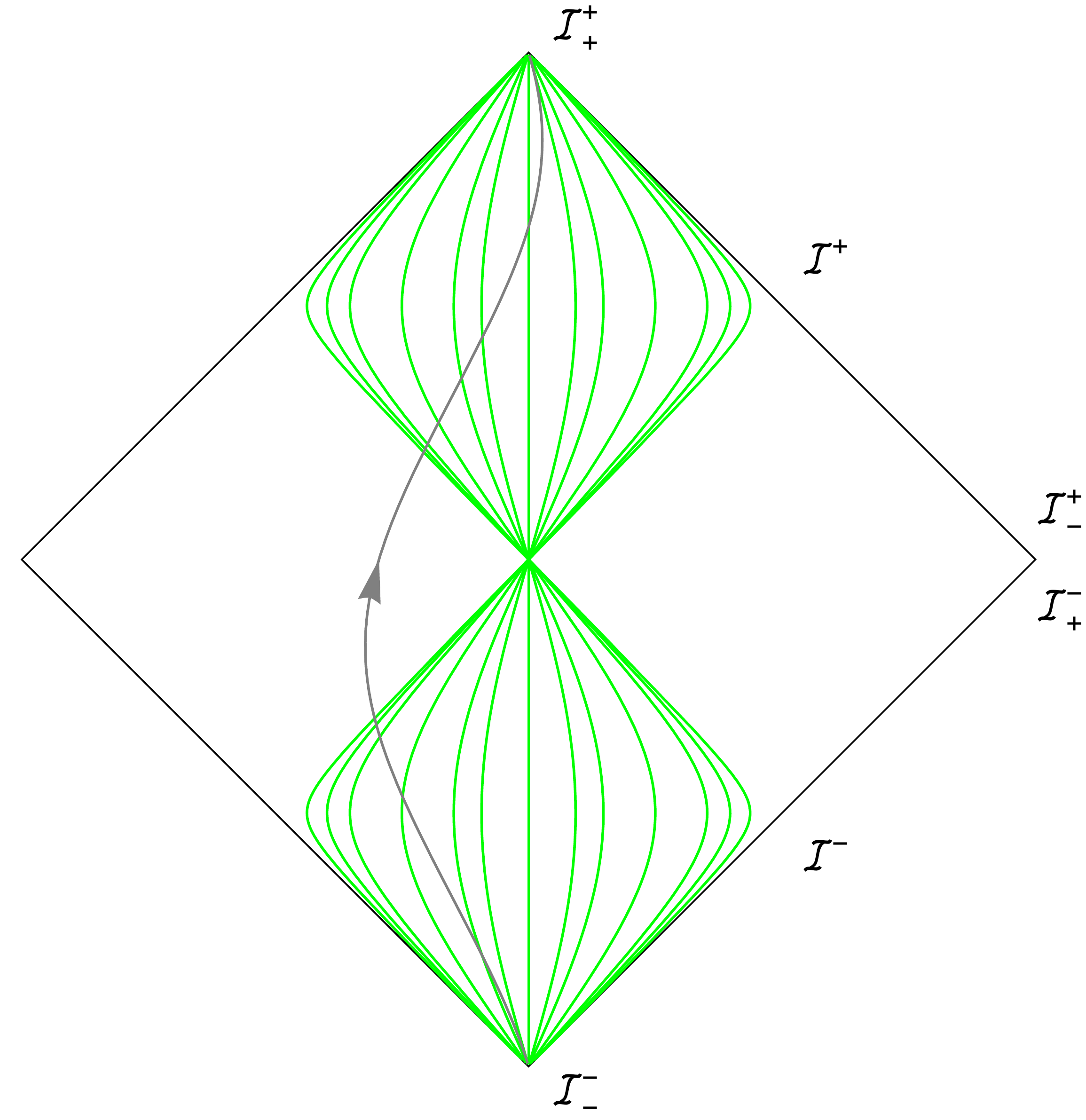}
\includegraphics[width=.49\textwidth]{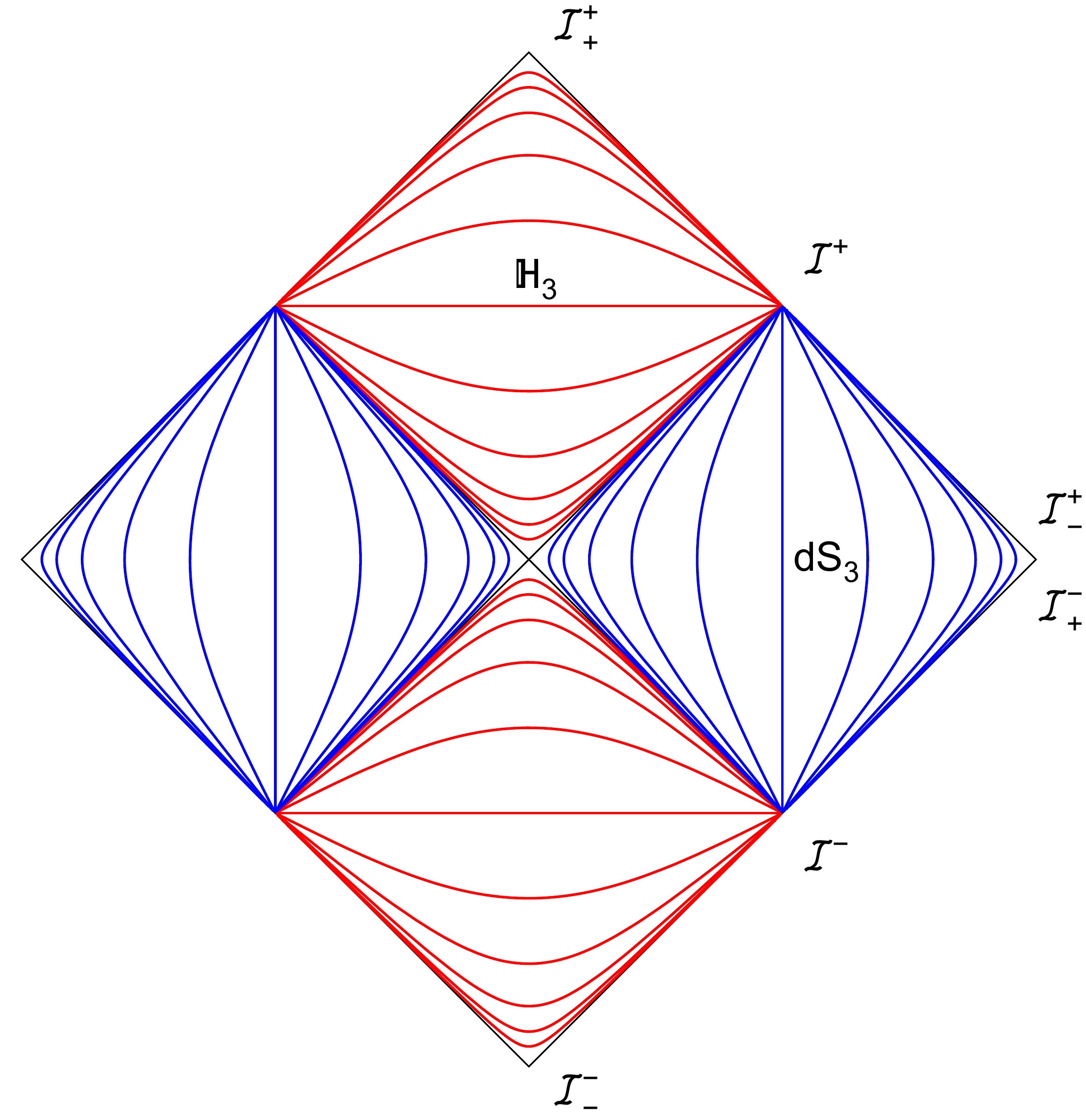}
\caption{Diagrams illustrating the causal structure of Minkowski spacetime, reproduced from \cite{Strominger:2017zoo}. Left: the green lines describe hypersurfaces of constant $\rho$, and the grey line is the world-line of a massive particle moving at a constant velocity. Right: hyperbolic slicing of Minkowski spacetime. The slices correspond to constant $\tau$ hypersurfaces, where for $\tau^2>0$ the resulting surface is the hyperbolic space $\mathbb{H}_3$ and for $\tau^2<0$ it is $\mathrm{dS}_3$.}
\label{fig:HyperbolicMinkowski}
\end{figure}

To study solutions to the Laplace equation \eqref{LaplaceEq}, we use the hyperbolic coordinates defined by
\begin{equation}
\tau =\sqrt{t^2 - r^2} = \sqrt{u^2 +2u r}\ ,
\qquad
\rho = \frac{r}{\sqrt{t^2 -r^2}}=\frac{r}{\sqrt{u^2 +2u r}}\ ,
\end{equation}
with the inverse relations given by
\begin{equation}
u = \tau \sqrt{1+ \rho^2} - \rho \tau\ ,
\qquad
r= \rho \tau
\ .
\end{equation}
The Minkowski metric then takes the form
\begin{equation}
ds^2 = -d\tau^2 +\tau^2 
\left(
\frac{d\rho^2}{1+\rho^2} +\rho^2 \gammaflat dz d\zb
\right)
\ .
\end{equation}
An illustrative diagram of the causal structure of Minkowski spacetime is given in Strominger's lecture notes \cite{Strominger:2017zoo},
	which we reproduce in figure \ref{fig:HyperbolicMinkowski}.

It was shown in \cite{Campiglia:2015kxa,Campiglia:2015lxa} that at $\tau \rightarrow \infty$ the only non-vanishing component of $\lambda_{\mu}$
	is $\lambda_{\tau}$,
\begin{equation}
\lim_{\tau \rightarrow \infty }\lambda_{\tau}(\tau,\rho,z,\zb) = \lamt_{\tau} (\rho,z,\zb)
\ ,
\end{equation}
In appendix \ref{GaugeModesApp} we study the two solutions for $\lamt_{\tau} (\rho,z,\zb)$.
At time-like infinity they asymptote to
\begin{equation}\label{lambdaSol}
\begin{aligned}
\lim_{\rho \rightarrow \infty } \lamt_{\tau} (\rho,z,\zb) 
& =
\alpha (z,\zb) \rho   \left( 1 + \dots  \right) 
+ \beta (z,\zb) \rho^{-3} \left( 1 + \dots   \right)   \ ,
\end{aligned}
\end{equation}
where the dots denote subleading terms in $1/\rho$.
The $\alpha$-series is leading and do not vanish at time-like infinity $\rho \rightarrow \infty$. It is a Large Gauge Transformation. The $\beta$-series is subleading and vanishes at time-like infinity.
We also show that in terms of the radiative data, the $\alpha$ and $\beta$ modes are given by
\begin{equation}
\begin{aligned}
 \alpha (z,\zb) &=  \left( \pa_z U_{\zb} + \pa_{\zb} U_z \right)_{\mI^+_+} \ , \\
\beta (z,\zb) &= i \left( \pa_z U_{\zb} - \pa_{\zb} U_z \right)_{\mI^+_+}
\ .
\end{aligned}
\end{equation}

The supertranslation charge (as well as the radiative data) is gauge-invariant to leading order in $r$ \cite{Campiglia:2015kxa,Avery:2015gxa,Avery:2015rga}.
At null infinity $\mathcal{\mI^+}$ it is given by
\begin{equation}
T(f) = \frac{1}{4\pi G} \int_{\mI^+_-}
d^2 z \gammaflat f(z,\zb) m_B
\ .
\end{equation}
Using the constraint equation \eqref{constraintEq} we can write
\begin{equation}
T(f) =\Tsoft (f) + T_\text{hard}(f),
\end{equation}
where we have the soft part, given by the boundary term
\begin{equation}\label{softPart}
\begin{aligned}
\Tsoft (f) 
&= \frac{1}{8 \pi G} \int  d^2z  \left[  \pa_z U_{\zb} +\pa_{\zb} U_{z} \right]  f(z,\zb),
\end{aligned}
\end{equation}
and the hard part
\begin{align}
\label{Thard}
T_\text{hard}(f)= \frac{1}{4\pi G} \int
du d^2 z  f(z,\zb)  \gammaflat T_{uu}\ .
\end{align}
The soft part of BMS supertranslations corresponds to the $\alpha$-mode.
The reason is that the graviton's zero-mode is precisely the pure gauge mode $\lambda_{\mu}$.
To isolate the BMS mode we therefore have to impose the following boundary conditions
\begin{equation}\label{bc}
\beta = i \left( \pa_z U_{\zb} - \pa_{\zb} U_z \right)_{\mI^+_{\pm}}
=
i \left(  D_z^2 C_{\zb\zb} - D_{\zb}^2 C_{zz}  \right)_{\mI^+_{\pm}} =0
\ .
\end{equation}
After imposing these boundary conditions, the Bondi news and the radiative data transform as
\begin{equation}
\begin{aligned}
\delta_f N_{zz} &= f \pa_u N_{zz} \\
\delta_f C_{zz} &= f \pa_u C_{zz} -2 D^2_z f
\end{aligned}
\end{equation}
under supertranslations, and the action of the BMS charge is described by the following Dirac (or Poisson) brackets
\begin{equation}
\begin{aligned}
\{ T(f) , C_{zz}  \} &= f \pa_u C_{zz} -2 D^2_z f\\
\{ T(f) , N_{zz}  \} &= f \pa_u N_{zz}
\ .
\end{aligned}
\end{equation}
Without imposing the boundary conditions \eqref{bc} the result of the Dirac brackets would be different. Imposing different boundary conditions will not change the Dirac brackets, but will fail to identify the BMS mode correctly (at leading order the $\beta$-mode does not contribute, but at subleading orders it will).


We would now like to express the generator of soft supertranslations in terms of the creation and annihilation operators.
Up to this point we have been using the Gupta-Bleuler quantization, but for the explicit computation below we will go further and adopt the canonical quantization in terms of the physical, transverse-traceless, components of the graviton
\begin{equation}
a_{\mu\nu} (k) = \sum_{r= \pm} \ep_{\mu\nu}^{r*} (k) a_r (k),
\end{equation}
where the momentum modes in the polarization basis obey the following commutation relations
\begin{equation}\label{aCommutations}
\left[
a_{r}(k), a^\dagger_{s}(k')
\right]
=\delta_{r s }  (2\omega_k)  (2\pi)^{3}  \delta^3 \left(\V{k}   -\V{k'} \right)
.
\end{equation}
The transverse-traceless components of the polarization tensor can be decomposed as follows
\begin{align}
\epsilon^\pm_\mn(k) = \epsilon^\pm_\mu(k)\epsilon^\pm_\nu(k),
\end{align}
and we will further use the following concrete realization for them
\begin{equation}\label{spin1pol}
\begin{aligned}
\ep^{-\mu} (k) &= \frac{1}{\sqrt{2}} \left( z,1,+i,-z \right),
\\
\ep^{+\mu} (k) &= \frac{1}{\sqrt{2}} \left( \zb,1,-i,-\zb \right).
\end{aligned}
\end{equation}
Let us also note that the four-momentum of the graviton, being massless, is given by
\begin{equation}
k^{\mu} = \frac{\omega_k}{1+z \zb} \left( 1+ z\zb,z+\zb, -i(z-\zb) , 1- z\zb \right).
\end{equation}


Using this and the plane wave expansion (for example see Appendix A of \cite{Gabai:2016kuf}),
	we write the radiative data as \cite{He:2014laa}
\begin{equation}
\begin{aligned}\label{CzzModes}
C_{zz} (u,z,\zb) &= \kappa \lim_{r\rightarrow \infty} \frac{1}{r} h_{zz}(r,u,z,\zb)\\
&= \kappa \lim_{r\rightarrow \infty} \frac{1}{r} \pa_z x^{\mu} \pa_z x^{\nu} h_{\mu\nu} \\
& = -\frac{i \kappa}{8\pi ^2 } \gammaflat
\int_0^\infty d \omega_k \left[
a_+(\omega_k \hat{\V{x}}_z) e^{-i \omega_k u}
-a^{\dagger}_-(\omega_k \hat{\V{x}}_z) e^{i \omega_k u}
\right].
\end{aligned}
\end{equation}
The soft supertranslations generator \eqref{softPart} can then be written as\footnote{We use that $\frac{1}{2\pi}\int du e^{-i \omega u}= \delta(\omega)$.
Note that since the $\omega$-integration is over half the real plane we have
\begin{equation*}
\int_0^{\infty} d \omega \, \delta(\omega) f(\omega) = \frac{1}{2} f(0)
\end{equation*}
}
\begin{equation}\label{TsoftModes}
\begin{aligned}
\Tsoft (f) 
&=  - \frac{1}{16 \pi G} \int du d^2z \left[
 N_{\zb}\,\! ^z D_z^2 f
+ N_{z}\,\! ^{\zb} D_{\zb}^2 f
\right] 
 \\
&=
  \lim_{\omega_k \rightarrow 0}\frac{ \omega_k }{4 \pi  \kappa}
  \int  d^2z \left[
\left(
a_+(\omega_k \hat{\V{x}}_z) 
+a^{\dagger}_-(\omega_k \hat{\V{x}}_z)
\right) D_{\zb}^2 f
+ \text{h.c.}
\right]. 
\end{aligned}
\end{equation}
In this form it is clear why $\Tsoft$ is, indeed, described by soft gravitons.

\section{Action of BMS supertranslation}\label{ActionOfBMS}

In this section we study the action of BMS supertranslations on single-particle states, as well as on the vacuum state, using the expressions obtained in section \ref{BMS}.

\subsection{Outgoing graviton}

Using the expression \eqref{TsoftModes} and the commutation relations \eqref{aCommutations}, the action of the supertranslation generator on an outgoing soft graviton at future null infinity is given by
\begin{equation}\label{BMSgraviton}
\left[ T(f) , a_{+/-} ({\bold k})  \right] =  \frac{8 \pi ^2}{ \kappa} \frac{1}{\gamma_{z_k \zb_k}} \delta(\omega_k)  D^2_{z/\zb} f.
\end{equation}
Since we take $a_{+/-} ({\bold k})$ to be soft, it is the soft part $\Tsoft(f)$ of $T(f)$ that contributes to \eqref{BMSgraviton}, and hence the delta function on the right hand side.

\subsection{Undressed massive particle}

The action of BMS supertranslations on an undressed massive particle has been studied in detail
	by Campiglia and Laddha \cite{Campiglia:2015kxa,Campiglia:2015lxa,Campiglia:2016efb}.
Here we briefly review this result.
For simplicity we take the particle to be a scalar, but to leading order the result will be the same for particles of any spin.

The retarded system of coordinates $(u,r,z,\zb)$ is useful to describe null infinity, and therefore more convenient when we discuss massless particles. However, massive particles reach null infinity only asymptotically (in the future), and to describe them it is more convenient to use the hyperbolic system of coordinates that we have introduced in subsection \ref{GaugeModes}.

The canonically quantized massive scalar field is given by \eqref{eqn:phi},
\begin{equation}\label{scalarModes}
\varphi(x) = \int \td{p} \left[ 
b(p) e^{ip \cdot x}
+b^{\dagger}(p) e^{-ip \cdot x}
\right].
\end{equation}
The creation and annihilation operators of the scalar particle obey the commutation relation
	\eqref{eqn:bcomm},
\begin{equation}
\left[ 
b(p) , b^{\dagger} (p')
\right]
=(2\pi)^3 (2\omega_p) \delta ^3 (\V{p}-\V{p}\,'),
\end{equation}
where $\omega_p^2 = |\V{p}|^2 +m^2$.
The phase factor is
\begin{equation}
x \cdot p = \tau \left(  \rho \, \hat{\V{x}} \cdot \V{p} - \omega_p \sqrt{1+\rho^2} \right).
\end{equation}
At large $\tau$ the integral in \eqref{scalarModes} is dominated by a saddle point at $\V{p} = m \rho \hat{\V{x}}$,
\begin{equation}
\lim_{\tau \rightarrow \infty} \varphi(x) =
\frac{\sqrt{m}}{2(2 \pi \tau)^{3/2}}
\left[
b (m\rho \hat{\V{x}}) e^{-im\tau}
+
b^{\dagger} (m\rho \hat{\V{x}})e^{im \tau}
\right],
\end{equation}
where the constant phase factors have been absorbed into the creation and annihilation operators.
Asymptotically, the scalar field transforms under BMS supertranslations as
\begin{equation}
\delta_f \varphi =\lamt_{\tau} (\rho,z,\zb)  \pa_{\tau} \varphi.
\end{equation}
The annihilation operator therefore transforms as
\begin{equation}
\delta_f b(p) =  -i m  \lamt_{\tau}(|\V{p}|/m , z ,\zb) \,  b(p),
\end{equation}
which is equivalent to the following commutation relation
\begin{equation}\label{Tb}
\begin{aligned}
\left[ T(f) , b(p) \right] 
&=
-m  \lamt_{\tau}(|\V{p}|/m , z ,\zb) b(p)\\
&=
-b(p)
\int \frac{d^2z}{4\pi} \sqrt{\gamma} 
\frac{m^4}{\left( \sqrt{m^2+  |\V{p}|^2} 
- \V{p} \cdot \hat{\V{x}}_{z}
 \right)^3}
 f(z,\zb).
\end{aligned}
\end{equation}

\subsection{Vacuum}\label{Vacuum}

BMS supertranslations give rise to a freedom in the definition of the vacuum.
We define the vacuum as the state that satisfies
\begin{equation}\label{vacuum}
 a(\omega \hat{\V{x}}) \ket{0} = 0
\ ,
\end{equation}
which applies, in particular, to a soft graviton annihilation operator.
Alternatively, the state $T(f) \ket{0}=\Tsoft(f) \ket{0}$, for any function $f(z,\zb)$, could serve as the zero energy state (note that the hard part inside $T(f)$ annihilates the vacuum).
Physically, this state differs from the original vacuum \eqref{vacuum} by the addition of a soft graviton.
In this section we show that these different choices are orthogonal to each another.
More explicitly, we show that acting with the generator of BMS supertranslations on the original vacuum \eqref{vacuum} creates a state which is orthogonal to any state constructed from the original vacuum
\begin{equation}\label{Orthogonal}
\bra{0} T(f) \hat{\Psi}^\text{out} e^{-R_f} \mS e^{R_f}  \hat{\Psi}^\text{in}  \ket{0} =0 \ .
\end{equation}
This implies that no physical process can transform the original vacuum into the new state generated by BMS supertranslations.
This is one of our main results in this paper.

We start by considering a scattering process with an emission of a single soft graviton.
The amplitude for this process is given by
\begin{equation}
\begin{aligned}
\Mksoft
	&= \braket{\text{out}|e^{-R_f}\mS e^{R_f}|\text{in}} \\
	&= \braket{\mathbf{k},r|\hat{\Psi}^\text{out}e^{-R_f}\mS e^{R_f}\hat{\Psi}^\text{in}|0}
\ ,
\end{aligned}
\end{equation}
where\footnote{We have used $\ep^{t} \cdot\ep^{-r} = \delta^{tr}$.}
\begin{equation}
\bra{\V{k},r} = \bra{0} a_r (k)=
 \epsilon^{r}_{\rho\sigma}(k) \bra{0} a^{\rho\sigma}(k) 
\end{equation}
is the soft graviton state with polarization $r$. The scalar operators are given by
\begin{equation}
	\hat{\Psi}^\text{in} \equiv \prod_{i\in\text{in}} b^\dagger(p_i)
	\qquad\text{and}\qquad
	\hat{\Psi}^\text{out} \equiv \prod_{j\in\text{out}} b(p_j),
\end{equation}
where ``in" and ``out" denote the set of incoming and outgoing scalar particles, respectively.
\begin{figure}[t]
\centering
\includegraphics[width=.19\textwidth]{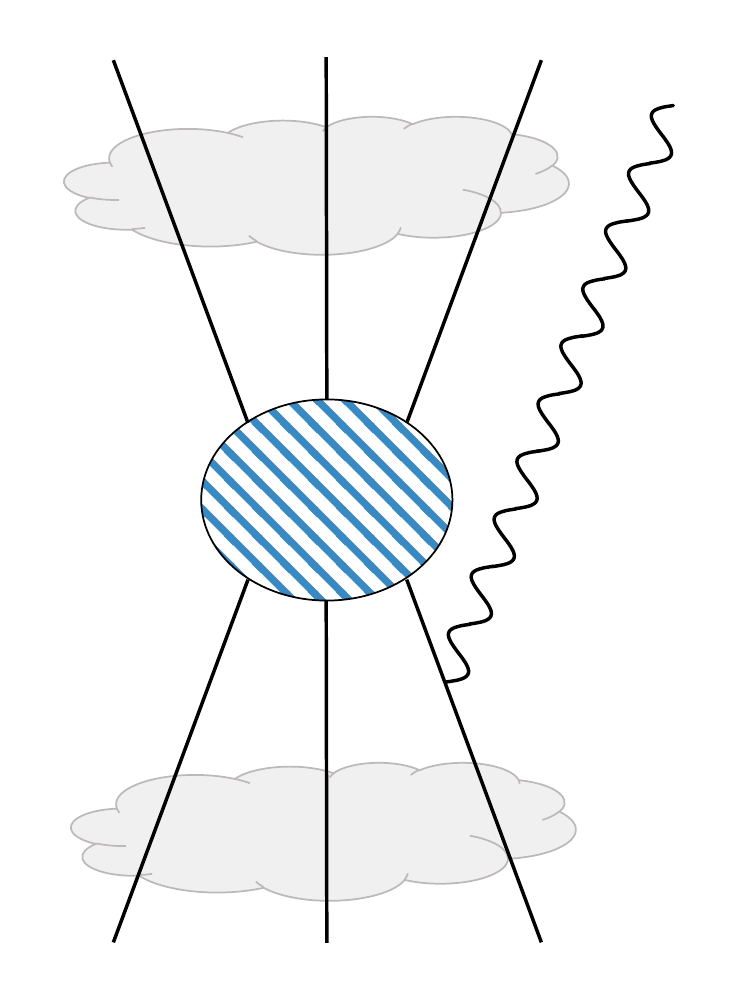}
\includegraphics[width=.19\textwidth]{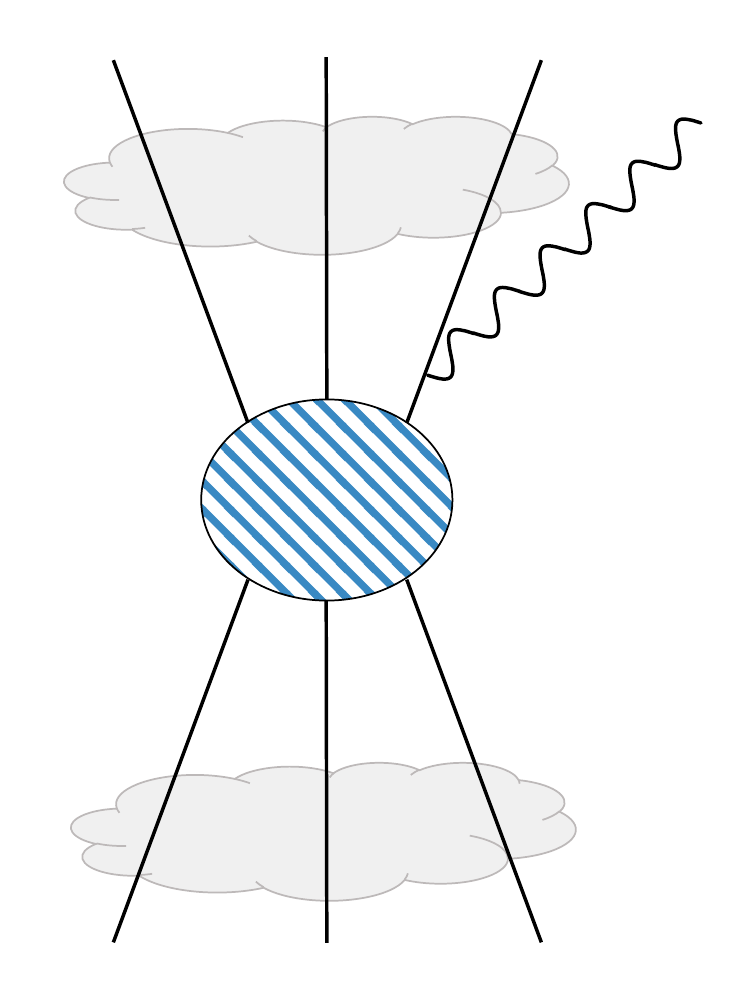}
\includegraphics[width=.19\textwidth]{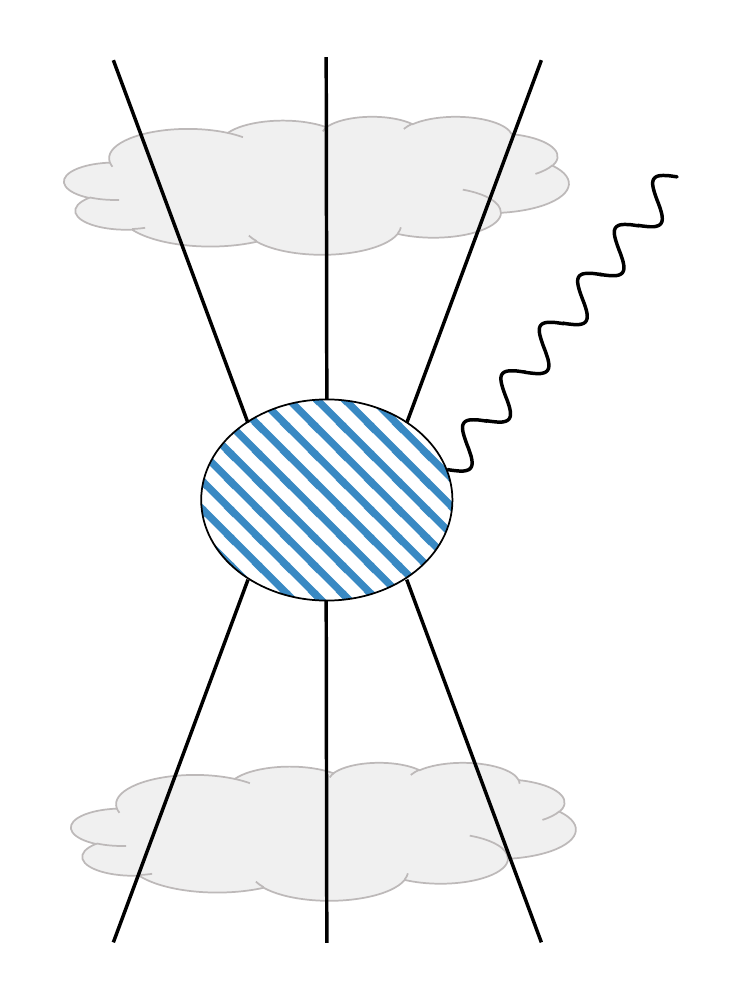}
\includegraphics[width=.19\textwidth]{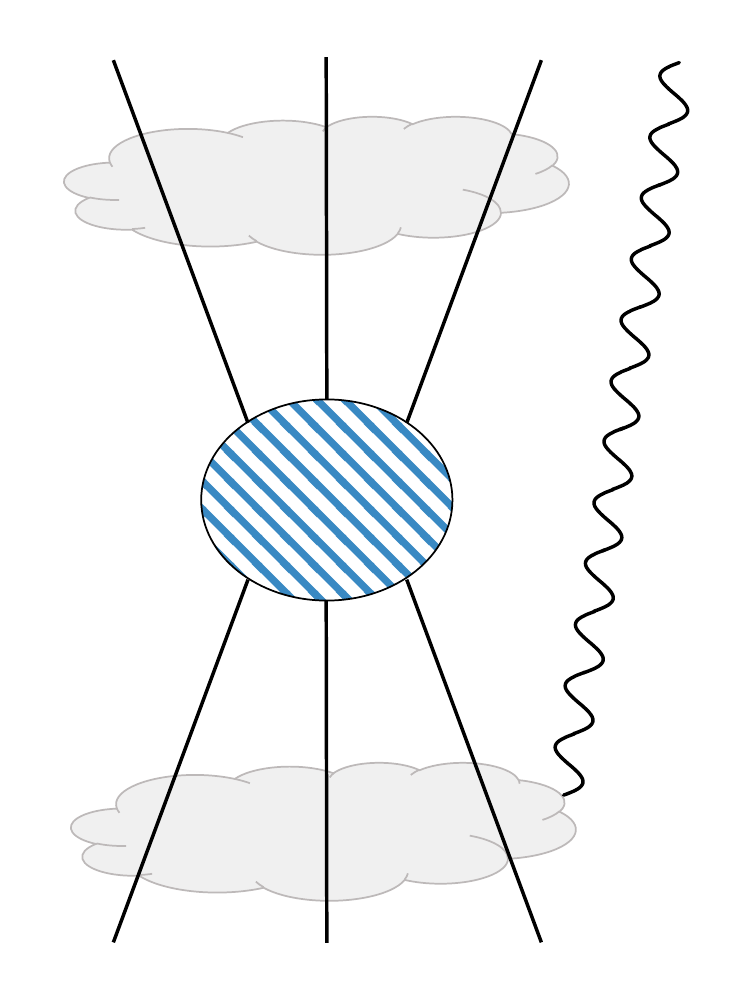}
\includegraphics[width=.19\textwidth]{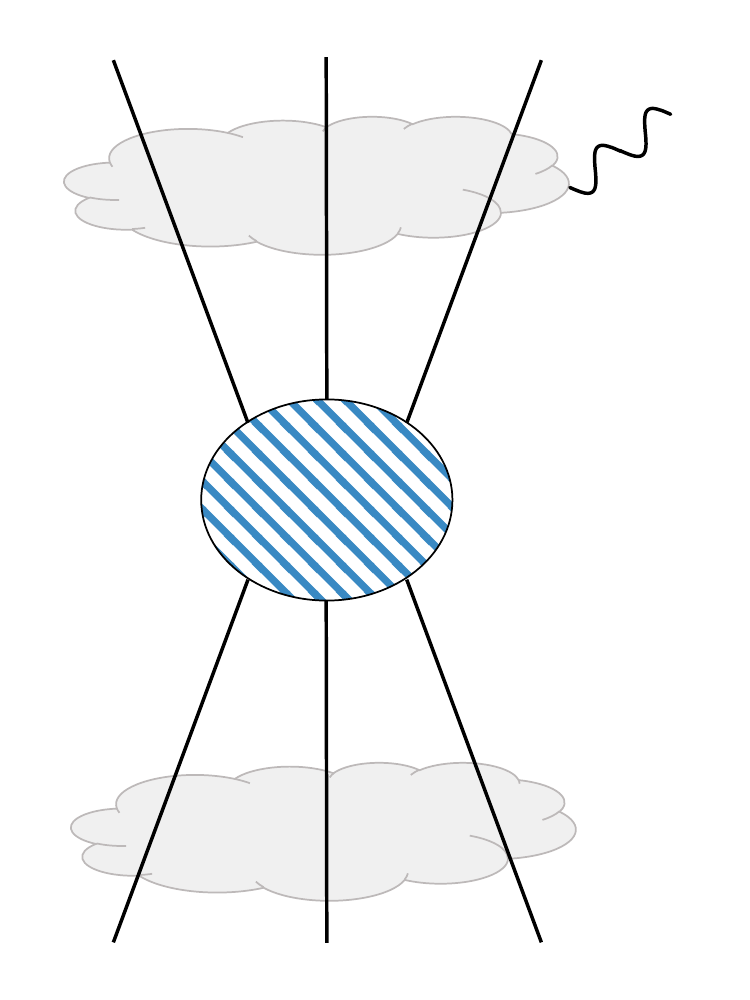}
\caption{Different ways to connect an external soft graviton to a scattering amplitude. The first two diagrams on the left represent a soft graviton that is connected to an external leg. The last two diagrams on the right represent a soft graviton that is connected to the gravitons' cloud. The diagram in the middle represents a soft graviton that is connected to an internal leg.}
\label{fig:singlegrav}
\end{figure}
The soft graviton can connect to a diagram in three different ways:
\begin{enumerate}[itemsep=0pt]
	\item Connect to an external scalar leg.
	\item Connect to the graviton cloud $e^{\pm R_f}$ (or equivalently $e^{\pm \Rfp}$).
	\item Connect to an internal leg.
\end{enumerate}
These three options are depicted in figure \ref{fig:singlegrav}.
Contractions of the last type are IR-convergent, and therefore will not contribute to the amplitude at leading order.
The cloud also dresses the soft graviton operator, but this dressing involves the scalar number operator $\rho(p)=b^\dagger(p)b(p)$
and will vanish by acting on the vacuum.

Consider the contraction of the first type.
By virtue of the soft theorem \cite{Weinberg:1965nx}, each such contraction contributes
\begin{align}
\label{eqn:soft_factor_form}
&\eta \,  \kappa \, \frac{p_\mu p_\nu}{2p\cdot k} \epsilon^{r}_\mn(k) \mM
\ ,
\end{align}
where $\eta=+1$ for an outgoing state and $\eta=-1$ for an incoming state, and
\begin{equation}
\mM\equiv\bra{0}\hat{\Psi}^\text{out}e^{-R_f}\mS e^{R_f}\hat{\Psi}^\text{in}\ket{0}
\end{equation}
is the amplitude without the soft graviton.
Let us briefly review the derivation of \eqref{eqn:soft_factor_form}.
Using the commutation relation \eqref{eqn:acomm}, we derive the momentum-space contraction rule to be
\begin{equation}\label{GravitonWavefunction}
\begin{aligned}
\contraction{}{\bra{\mathbf{k},r}}{}{h}
\bra{\mathbf{k},r}h_{\mu\nu}
	&= \epsilon^{r,\rho\sigma}(k)
		\bra{0}\int\td{k'}
		\left[a_\rs(k),a^{\dagger}_\mn(k')\right] \\
	&= \frac{1}{2} \epsilon^{r,\rs}(k) I_\rsmn \bra{0}\\
	&= \epsilon^{r}_\mn(k)\bra{0},
\end{aligned}
\end{equation}
where we used \eqref{TTcomponents} in the last line.
One may consider this to be the external ``wavefunction" of a graviton with polarization $r$.
Next, we observe that the insertion of a soft graviton to an external leg with momentum $p$ adds a scalar propagator
\begin{align}
\frac{-i}{(p\pm k)^2 + m^2}
	\quad \xrightarrow{k\rightarrow 0} \quad
	\mp\frac{i}{2p\cdot k}
\end{align}
and scalar-scalar-graviton vertex
\begin{align}\label{vertex}
\frac{i\kappa}{2} \left(
		p_\mu (p\pm k)_\nu + p_\nu (p\pm k)_\mu - \frac{1}{2}\eta_{\mu\nu} \left[p\cdot(p\pm k)+m^2\right]
	\right)
	\quad \xrightarrow{k\rightarrow 0} \quad
	i\kappa p_\mu p_\nu\ ,
\end{align}
where the upper (lower) sign is for an outgoing (incoming) state.
Putting \eqref{GravitonWavefunction}-\eqref{vertex} together, we recover the result \eqref{eqn:soft_factor_form} in the soft limit.

Next, we will study the soft gravitons' contractions of the second type, i.e. to the clouds of gravitons.
The incoming and outgoing asymptotic states can be written in terms of single particle dressed states:
\begin{align}
e^{R_f}  \hat{\Psi}^\text{in} \ket{0}  &=\left[\prod_{i\in\text{in}} b^\dagger(p_i) e^{R_f(p_i)} \right] \ket{0}
\end{align}
and
\begin{align}
\bra{0}\hat{\Psi}^\text{out}e^{-R_f} &= \bra{0} \left[\prod_{j\in\text{out}} e^{-R_f(p_j)}b(p_j)\right].
\end{align}
The contraction of the soft graviton with the cloud gives
\begin{equation}\label{contractionExtCloud}
\begin{aligned}
\contraction{}{\bra{\mathbf{k},r}}{}{e^{\pm \Rfp}}
\bra{\mathbf{k},r}e^{\pm \Rfp}
&= 
\contraction{\pm}{\bra{\mathbf{k},r}}{}{R_f}
\pm\bra{\mathbf{k},r} \Rfp e^{\pm \Rfp} \\
&= \pm \epsilon^{r,\rs}(k)\bra{0}
	\frac{\kappa}{2}\int\td{k'}f^{\mu\nu}(k',p)
	\left[a_\rs(k),a^{\dagger}_\mn(k')\right]e^{\pm R_f(p)} \\
&= \pm \frac{\kappa}{4}f^{\mu\nu}(p,k)I_\mnrs\epsilon^{r,\rs}(k)
	\bra{0}e^{\pm R_f(p)} \\
&= \pm \frac{\kappa}{2}f^\mn(p,k)\epsilon^{r}_\mn(k)
	\bra{0}e^{\pm R_f(p)} \ ,
\end{aligned}
\end{equation}
where the upper (lower) sign is for incoming (outgoing) particles.

Summing over all the diagrams and taking the soft limit results in
\begin{equation}\label{softDiagram}
\begin{aligned}
\lim_{\omega_k \rightarrow 0}
&\omega_k \, \Mksoft
=
\lim_{\omega_k \rightarrow 0}
\omega_k \,
\\& \times
\frac{\kappa}{2}\left[\sum _{j\in\text{out}} \frac{p^\mu_j p^\nu_j}{p_j\cdot k}\Big(1-\phi(p_j,k)\Big)
-\sum _{i\in\text{in}} \frac{p_i^\mu p_i^\nu}{p_i\cdot k}\Big(1-\phi(p_i,k)\Big)
\right. \\ & \left.
-\frac{1}{\wk}\left(\sum _{j \in \text{out}} c'^{\mu\nu}(p_j,k)\phi(p_j,k) 
-\sum _{i \in \text{in}} c^{\mu\nu}(p_i,k)\phi(p_i,k) 
 \right)\right]
\epsilon^{r}_\mn(k)\mM,
\end{aligned}
\end{equation}
where the c-matrix $c_\mn$ ($c'_\mn$) was used to construct the incoming (outgoing) state. 
By definition $\lim_{\omega_k \rightarrow 0}\phi(p,k)=1$,
	and the first two sums inside the square brackets of \eqref{softDiagram} vanish.
The second line becomes
\begin{align}
&\lim_{\wk\rightarrow 0}
\frac{\kappa}{2}
\left[
	\sum _{j \in \text{out}} c'^{\mu\nu}(p_j,k)
	-\sum _{i \in \text{in}} c^{\mu\nu}(p_i,k)
\right]\epsilon^r_\mn(k) \mM.
\end{align}
With the parametrization \eqref{cmatrix}, we observe that using the same $q$ for incoming and outgoing states reduces this to
\begin{align}
&\lim_{\wk\rightarrow 0}
\frac{\kappa}{2}
\left[
	\frac{\wk}{k\cdot q}
	\left(
		\frac{k \cdot p_\text{tot}}{k\cdot q}q^\mu q^\nu
		- 2q^\mu p^\nu_\text{tot}
	\right)
\right]\epsilon^r_\mn(k) \mM=0,
\end{align}
since
\begin{align}
p_\text{tot}\equiv\sum_{j\in\text{out}}p_j - \sum_{i\in\text{in}}p_i=0
\end{align}
by energy-momentum conservation.
For the general case where $c'_\mn\neq c_\mn$, we show at the end of \cref{app_ir} that processes with non-zero amplitudes can only occur
	between states that satisfy
\begin{align}\label{SumEqual}
\sum _{j \in \text{out}} c'^{\mu\nu}(p_j,k)
=\sum _{i \in \text{in}} c^{\mu\nu}(p_i,k).
\end{align}
Therefore, we conclude that
\begin{equation}\label{softAmp}
\lim_{\omega_k \rightarrow 0} \omega_k \Mksoft= 0\ .
\end{equation}
Since the creation operators annihilate the Bra vacuum, the action of the soft part of the BMS supertranslations \eqref{TsoftModes} is given by
\begin{equation}\label{superSoft}
\left< 0|T(f) \right. =
  \lim_{\omega_k \rightarrow 0}\frac{ \omega_k }{4 \pi  \kappa}
  \int  d^2z \bra{0} \left[
a_-(\omega_k \hat{x}) D_z^2 f
+a_+ (\omega_k \hat{x}) D_{\zb}^2 f
\right] .
\end{equation}
The soft limit of the amplitude, equation \eqref{softAmp}, together with \eqref{superSoft}, then implies the identity \eqref{Orthogonal}. Namely, the original vacuum state $\ket{0}$ and and the new state $\Tsoft(f) \ket{0}$ are orthogonal.

\section{BMS supertranslation of asymptotic states}\label{DressedParticles}

We are now in a position to compute BMS supertranslations of a physical asymptotic state. For simplicity we consider a single particle state with momentum $p$ dressed with a graviton cloud. The action of the supertranslation generator on the physical asymptotic state can be decomposed into the following three pieces
\begin{equation}\label{dressedParticle}
\begin{aligned}
\bra{0} e^{-\Rfp} \, b(p) \,  T(f) 
&=
\bra{0}  T(f) \, e^{-\Rfp} \, b(p)\\
& 
+\bra{0} \left[ e^{-\Rfp} \ , T(f) \right]   b(p) 
-\bra{0} e^{-\Rfp} \, \left[T(f), b(p) \right] .
\end{aligned}
\end{equation}
The first term in \eqref{dressedParticle} is the action of BMS supertranslation on the vacuum.
It will vanish when contracted with an incoming (ket) state, by the result of previous section.
The second and third terms are the actions of BMS on the graviton cloud and on the massive particle, respectively.

Let us first compute the commutator of $\Rfp$ and $T(f)$,
\begin{equation}
\begin{aligned}
\left[ R_f(p) , T(f) \right]  &=
\frac{\kappa}{2} 
\int \td{k}
\left(
f^{\mu\nu}(p,k) \ep_{\mu\nu}^r \left[ a_{r}^{\dagger}(k) , T(f) \right]
-\text{h.c.}
\right).
\end{aligned}
\end{equation}
Using \eqref{BMSgraviton} we arrive at
\begin{equation}
\begin{aligned}
\left[ R_f(p) , T(f) \right]  
&= 
- 4 \pi ^2 
\int 
 \frac{\td{k}}{\gammaflatk}
\delta(\omega_k)
\left[
 f^{\mu\nu}(p,k)
\left(  
\ep_{\mu\nu}^-  D^2_{z}f
+ \ep_{\mu\nu}^+ D^2_{\zb}f
\right)
+\text{h.c.}
\right]\\
&=
- \pi ^2
\int  \frac{   d^2 z}{(2\pi)^3} 
\omegak 
\left[
 f^{\mu\nu}(p,k)
\left(  
\ep_{\mu\nu}^-  D^2_{z}f
+ \ep_{\mu\nu}^+ D^2_{\zb}f
\right)
+\text{h.c.}
\right].
\end{aligned}
\end{equation}
Defining $\hat{k}_{z,\zb}\equiv ( 1, \hat{{\bold k}}_{z,\zb} )$, the last expression takes the form
\begin{equation}
\begin{aligned}
\left[ R_f(p) , T(f) \right]  
  &= 
- \pi ^2
\int  \frac{   d^2 z}{(2\pi)^3} 
\left(  \frac{p^{\mu}p^{\nu}}{\hat{k}_{z,\zb}\cdot p} +c^{\mu\nu }  \right) 
\left[
\left(  
\ep_{\mu\nu}^- (\hat{k}_{z,\zb})   D^2_{z}f
+ \ep_{\mu\nu}^+(\hat{k}_{z,\zb})  D^2_{\zb}f
\right)
+\text{h.c.}
\right],
\end{aligned}
\end{equation}
where according to our convention the delta function yielded half the value of the integrand at $0$.
Since $\ep_{\mu\nu}^{r*} = \ep_{\mu\nu}^{-r} $, we arrive at
\begin{equation}
\begin{aligned}
\left[ R_f(p) , T(f) \right] 
  &= 
-2 \pi ^2
\int  \frac{   d^2 z}{(2\pi)^3} 
\left(  \frac{p^{\mu}p^{\nu}}{\hat{k}_{z,\zb}\cdot p} +c^{\mu\nu }  \right) 
\left(  
\ep_{\mu\nu}^- (\hat{k}_{z,\zb})   D^2_{z}f
+ \ep_{\mu\nu}^+(\hat{k}_{z,\zb})  D^2_{\zb}f
\right).
\end{aligned}
\end{equation}
Integrating by parts, we then have\footnote{Derivatives with upper indices are defined as usual by $\pazbU= \gamma^{\zb z}\paz$ and $\pazU= \gamma^{z\zb}\pazb$.}
\begin{equation}
\begin{aligned}
\left[ R_f(p) ,  T(f) \right] 
  &= 
-\int  \frac{   d^2 z}{4\pi} 
\left[  
\paz \pazbU \left( \gammaflat \frac{p^{\mu}p^{\nu} \ep_{\mu\nu}^- }{\hat{k}_{z,\zb}\cdot p}   \right)   
+\pazb \pazU \left( \gammaflat  \frac{p^{\mu}p^{\nu}\ep_{\mu\nu}^+}{\hat{k}_{z,\zb}\cdot p}   \right)  
+C(p,z,\zb) 
\right]f\ ,
\end{aligned}
\end{equation}
where we have defined
\begin{equation}
C(p,z,\zb) \equiv
\paz \pazbU ( \gammaflat c^{\mu\nu } \ep_{\mu\nu}^-   )
+\pazb \pazU ( \gammaflat c^{\mu\nu }  \ep_{\mu\nu}^+ ).
\end{equation}
With $\ep^{\pm}_\mn=\ep^{\pm}_{\mu}\ep^{\pm}_{\nu}$, we get
\begin{equation}
\begin{aligned}
\left[ R_f(p) ,  T(f) \right] 
  &= 
-\int  \frac{   d^2 z}{4\pi} 
\left[  
\paz \pazbU    \left(    \gammaflat  \frac{ (p \cdot \ep^-)^2 }{\hat{k}_{z,\zb}\cdot p}   \right)   
+\pazb \pazU    \left(    \gammaflat  \frac{(p \cdot \ep^+)^2}{\hat{k}_{z,\zb}\cdot p}   \right)  
+C(p,z,\zb) 
\right]f\ .
\end{aligned}
\end{equation}
An explicit calculation shows that
\begin{equation}
\paz \pazbU    \left(    \gammaflat  \frac{ (p \cdot \ep^-)^2 }{\hat{k}_{z,\zb}\cdot p}   \right)   
=\pazb \pazU    \left(    \gammaflat  \frac{(p \cdot \ep^+)^2}{\hat{k}_{z,\zb}\cdot p}   \right)  
=
\frac{1}{2} \gammaflat \frac{p^4}{(p\cdot \hat{k}_{z,\zb} )^3}\ .
\end{equation}
We therefore end up with
\begin{equation}\label{eqn:RT}
\begin{aligned}
\left[ R_f(p) ,  T(f) \right] 
  &= 
-\int  \frac{   d^2 z}{4\pi} 
\left[  
\gammaflat \frac{p^4}{(p\cdot \hat{k}_{z,\zb} )^3}
+C(p,z,\zb) 
\right]f\ .
\end{aligned}
\end{equation}
The first contribution,
\begin{equation}
 \frac{p^4}{(p\cdot \hat{k}_{z,\zb} )^3} =  \frac{m^4}{\left({\bold p} \cdot \hat{{\bold k}}_{z,\zb} -\sqrt{m^2 +|\V{p}|^2} \right)^3}
 \ ,
\end{equation}
is the Aichelburg-Sexl gravitational field of a massive particle \cite{Aichelburg:1970dh}. This is the gravitational analogue of the Lienard-Wiechert electromagnetic radiation field of a moving charged particle.

We now see that the first term in the second line of \eqref{dressedParticle} is equal to
\begin{equation}
\bra{0} e^{-\Rfp} \left[ -\Rfp \, , T(f) \right] b(p)
\ ,
\end{equation}
since $\left[ \Rfp \ , T(f) \right]$ is a c-number.
We therefore have
\begin{equation}\label{eqn:RTbTvanish}
\begin{aligned}
\bra{0} e^{-\Rfp} \, b(p) \,  T(f) 
=&
\bra{0}  T(f) \, e^{-\Rfp} \, b(p)\\
&\quad-\bra{0} e^{-\Rfp}
\Big\lbrace  \left[ \Rfp \, , T(f) \right] b(p)+\left[ T(f), b(p)\right]
\Big\rbrace.
\end{aligned}
\end{equation}
From \eqref{Tb} and \eqref{eqn:RT}, 
	we observe that the BMS of the bare particle and the momentum dependent part of the BMS of the graviton
	cloud exactly cancel each other\footnote{Note that $\sqrt{\gamma}=\gammaflat$.}.
Finally, the outgoing BMS charge between the two physical asymptotic states is then given by
\begin{equation}\label{BMScharge}
\frac{\bra{0}\hat{\Psi}_\text{as}^\text{out} \Tsoft \mS\hat{\Psi}_\text{as}^\text{in}\ket{0}}
	{\bra{0}\hat{\Psi}_\text{as}^\text{out}\mS\hat{\Psi}_\text{as}^\text{in}\ket{0}}
=
-\sum_{j \in \text{out}}
\int
\frac{d^2z}{4\pi} C(p_j,z,\zb) f(z,\zb).
\end{equation}
Similarly, one can also construct the BMS charge of an incoming physical asymptotic state.
The BMS charge \eqref{BMScharge} parameterizes the asymptotic state and is conserved as long as BMS supertranslation is a symmetry of the system, in line with the discussion in section \ref{AsymStates}.

To better understand the meaning of the BMS charge and the implications of the BMS symmetry, we end this section by looking at a BMS eigenstate defined as
\begin{equation}
\bra{\Omega_{\Lambda} } T(f)  \equiv \int \frac{d^2 z}{4 \pi} \Lambda(z,\zb) f(z,\zb)\bra{\Omega_{\Lambda} },
\end{equation}
and which is related to the vacuum by
\begin{equation}
\bra{0} = \int \mD [\Lambda] e^{-\frac{1}{2} \Lambda^2} \bra{\Omega_{\Lambda}},
 \qquad \text{where} \qquad
\Lambda^2 = \int \frac{d^2 z}{4 \pi } \Lambda^2 (z,\zb)
\ .
\end{equation}
in a similar fashion to the case of QED \cite{Gabai:2016kuf}.
The asymptotic states built from these eigenstates are also eigenstates of BMS transformations
\begin{equation}
\bra{\Omega_{\Lambda} }  e^{-\Rfp} b(p)  T(f)   =
\bra{\Omega_{\Lambda} }  e^{-\Rfp} b(p)   \int \frac{d^2 z}{4 \pi}\left[ \Lambda(z,\zb)- C(p,z,\zb) \right] f(z,\zb),
\end{equation}
and similarly their BMS charge is given by
\begin{equation}
\frac{\bra{\Omega_{\Lambda} }  \hat{\Psi}_\text{as}^\text{out} \Tsoft \mS\hat{\Psi}_\text{as}^\text{in}\ket{\Omega_{\Lambda} }}
	{\bra{\Omega_{\Lambda} }\hat{\Psi}_\text{as}^\text{out}\mS\hat{\Psi}_\text{as}^\text{in}\ket{\Omega_{\Lambda} }}
=
   \int \frac{d^2 z}{4 \pi}\left[ \Lambda(z,\zb)- \sum_{j \in \text{out}} C(p_j ,z,\zb) \right] f(z,\zb)
\end{equation}
The state $\ket{\Omega_{\Lambda}}$ belongs to a superselection sector which is characterized by its BMS charge.
We can now study the transition amplitude between two different BMS eigenstates by computing the expectation value of the following commutator
\begin{align}\label{AsympStates}
\begin{split}
\bra{\Omega_{\Lambda_1} } & \hat{\Psi}_\text{as}^\text{out}
		[T(f),\mS]
		\hat{\Psi}_\text{as}^\text{in}\ket{\Omega_{\Lambda_2} }
	\\&=
		\int \frac{d^2 z}{4 \pi}\left[ \Lambda_1(z,\zb)- \Lambda_2(z,\zb) \right] f(z,\zb)
		\bra{\Omega_{\Lambda_1} }  \hat{\Psi}_\text{as}^\text{out}
		\mS
		\hat{\Psi}_\text{as}^\text{in}\ket{\Omega_{\Lambda_2} },
\end{split}
\end{align}
where we used \eqref{SumEqual} to remove the terms involving $C(z,\zb)$.
The left hand side of equation \eqref{AsympStates} is the difference between the total incoming and outgoing BMS charges, which is zero by the conservation law of the symmetry.
The right hand side will vanish when either
\begin{equation}
\Lambda_1= \Lambda_2
\end{equation}
or
\begin{equation}
\bra{\Omega_{\Lambda_1} }  \hat{\Psi}_\text{as}^\text{out}\mS\hat{\Psi}_\text{as}^\text{in}\ket{\Omega_{\Lambda_2} } = 0
\qquad
\text{for}
\qquad
\Lambda_1  \neq  \Lambda_2
\end{equation}
We therefore conclude that BMS symmetry implies that the amplitude for transition between different superselection sectors is zero, once the contribution of the FK clouds is taken into account.

\section{Discussion}\label{conclusions}

In this paper we have studied the effect of BMS supertranslations on physical asymptotic states in perturbative quantum gravity. These states were constructed in \cite{Ware:2013zja} using the method of Kulish and Faddeev for QED \cite{Kulish:1970ut} by dressing the Fock states with a cloud of soft gravitons.
BMS supertranslations, in turn, give rise to a freedom in the definition of the vacuum. By acting with the BMS generator on the vacuum one generates a different state which could equally serve as the zero energy state. Therefore there exists a family of states generated by the action of BMS supertranslations on the vacuum.
This is a continuous family (or a moduli) which is parameterized by the BMS transformation parameter. 

Let us summarize our main results.
First, we have shown that all the states in this family are orthogonal to each other once we take into account the contribution of the FK clouds. In other words, the amplitude for transition between any two states in this family is zero for any physical process.
Second, we have computed the BMS charge of a physical asymptotic state. The BMS charge is conserved if BMS supertranslation is a symmetry of the system.
It characterizes the superselection sector to which the state belongs and the conservation law implies that there is no transition between different superselection sectors.

We would now like to make a comment about the role of zero-momentum modes in our computation.
The boundary condition that is imposed on the gauge modes \eqref{bc} alters the canonical commutation relations \eqref{aCommutations}.
It implies that the two polarization modes, at zero momentum, are related by
\begin{equation}
D_z^2 a_-(0) = D_{\zb}^2 a_+(0)\ .
\end{equation}
and do not affect non-zero momentum modes.
This is not surprising, since at zero momentum the two polarization modes are indistinguishable. 
However, the zero-momentum modes do not enter into our computations.
The reason is that all the external graviton states that we have considered are soft, but nevertheless of non-zero momentum, and therefore do not contract with strictly zero momentum modes.
Therefore it was safe for us to use the canonical commutation relations.
In \cite{Gabai:2016kuf}, for example, the commutation relations of the zero-momentum modes were used and therefore the authors had to correct the results by a factor of 2.

We end with a couple of future directions that we would like to pursue.
In this paper we have studied the soft dynamics in the asymptotic region of Minkowski spacetime.
It will be very interesting to use the same methods to study the soft dynamics near the black hole horizon.
In particular, we would like to understand the relation to the works of \cite{Mirbabayi:2016axw,Bousso:2017dny,Bousso:2017rsx}, where the authors have studied the effect of BMS transformations on the black hole soft hair, and to possibly extend their results.
Another interesting direction would be to extend our analysis to subleading orders in the soft approximation (see \cite{Laddha:2017ygw,Chakrabarti:2017ltl} for recent works on the subject).

\acknowledgments

We would like to thank Steven Avery, David Garfinkle, Daniel Grumiller, Carlos Hoyos, Andrew Strominger and Alexander Zhiboedov for useful discussions.
S.C. acknowledges a fellowship from the Samsung Foundation of Culture.

\appendix

\section{Convergence constraints}
\label{app_c}
Starting from the interaction term, one can show \cite{Ware:2013zja} that the graviton cloud operator
	is of the form $e^{R(t)}$, where
\begin{align}
R(t) &=
\label{DefOfRt2}
	\frac{\kappa}{2}
	\int \td{p}\td{k}
	\rho(p)
	\frac{p^\mu p^\nu}{p\cdot k}
	\left(
		a^\dagger_{\mu\nu}(k)e^{-i\frac{p\cdot k}{\omegap}t}
		- a_{\mu\nu}(k)e^{i\frac{p\cdot k}{\omegap} t}
	\right).
\end{align}
We used the shorthand notation \eqref{tilde_shorthand},
and $\rho(p)=b^\dagger(p)b(p)$ is the number operator of the scalar particle.
$e^{R(t)}$ maps the Fock space $\hf$ to the Faddeev-Kulish asymptotic space $\has$, i.e.
\begin{align}
e^{R(t)}\hf=\has\ .
\end{align}
An operator of the form $e^{R_f}$, where $R_f$ is given by
\begin{align}
\label{DefOfRf}
R_f =
	\frac{\kappa}{2}
	\int\td{p}\td{k}
	\rho(p)
	\left(
		f^{\mu\nu*}a^\dagger_{\mu\nu}
		- f^{\mu\nu}a_{\mu\nu}
	\right),
\end{align}
can be constructed such that $e^{R_f}$ also yields the Faddeev-Kulish asymptotic space:
\begin{align}
	\has = e^{R_f}\hf\ .
\end{align}
We wish to identify the constraints on $f^{\mu\nu}$ that allows the operator $e^{R_f}$ to have this property.
To this end, let us use the Baker-Campbell-Hausdorff (BCH) formula to decompose $e^{R(t)}$ as
\begin{align}
e^{R(t)}
	&= e^{R_f}e^{R(t)-R_f}e^{-\frac{1}{2}[R_f,R(t)]}.
\end{align}
Demanding that $e^{R(t)-R_f}$ and $e^{-\frac{1}{2}[R_f,R(t)]}$
	be unitary operators in the Fock space will yield the desired property,
\begin{align}
\has
	&=e^{R(t)}\hf
	= e^{R_f}e^{R(t)-R_f}e^{-\frac{1}{2}[R_f,R(t)]}\hf
	= e^{R_f}\hf\ .
\end{align}

Let us start with $e^{-[R_f,R(t)]/2}$.
The definitions \eqref{DefOfRt2} and \eqref{DefOfRf} tell us that both $R_f$ and $R(t)$ are anti-Hermitian.
Since the commutator of two anti-Hermitian operators is itself anti-Hermitian, 
	$e^{-[R_f,R(t)]/2}$ is a unitary operator (up to normalization) as long as the commutator converges.
By direct calculation, we obtain
\begin{align}
\label{RfRtMain}
\begin{split}
\left[R_f,R(t)\right]
	=& \frac{\kappa^2}{8}\int
		\td{p_1}\td{p_2}\td{k}
		\rho(p_1)\rho(p_2)
		\\&\times I_\mnrs
		\Big[
			f^{\mu\nu*}(p_1,k)\mP^{\rho\sigma}(p_2,k)
			-f^{\mu\nu}(p_1,k)\mP^{\rho\sigma*}(p_2,k)
		\Big],
\end{split}
\end{align}
where
\begin{align}
\mP^\mn(p,k)\equiv\frac{p^\mu p^\nu}{p\cdot k}e^{i\frac{p\cdot k}{\omegap} t}.
\end{align}
This commutator involves the $k$-integral
\begin{align}
\begin{split}
	&\int \frac{d^3k}{\wk}
	\frac{p^\rho_2 p^\sigma_2}{p_2 \cdot k}
	\phi(p_1,k)
	\left\{
		\left(\frac{p^\mu_1 p^\nu_1}{p_1 \cdot k}+\frac{c^{\mu\nu*}}{\wk}\right)
			e^{i\frac{p_2\cdot k}{\omega_{p_2}}t}
		- \left(\frac{p^\mu_1 p^\nu_1}{p_1 \cdot k}+\frac{c^{\mu\nu}}{\wk}\right)
			e^{-i\frac{p_2\cdot k}{\omega_{p_2}}t}
	\right\},
\end{split}
\end{align}
which has IR divergence if the leading term of $c_\mn$ in $k$ has non-zero imaginary part.
Therefore, the unitarity of $e^{-\frac{1}{2}[R_f,R(t)]}$ demands
\begin{align}
\label{c_real}
c^*_\mn(p,k)-c_\mn(p,k)=O(k).
\end{align}
The subleading terms of $c_\mn$ does not contribute to the commutator \eqref{RfRtMain};
the asymptotic time $t$ is taken to be very large, i.e. $|t|\rightarrow \infty$,
	and by virtue of the Riemann-Lebesgue lemma, the only contribution comes from small $k$.

Next, we consider $e^{R(t)-R_f}$. Using the BCH formula to write this in a normal-ordered form, we obtain
\begin{align}
\begin{split}
\label{eqn:e_diff}
e^{R(t)-R_f}
&=
	\exp\left\lbrace
		-\frac{\kappa^2}{16}
		\int\td{p_1}\td{p_2}\td{k}\rho(p_1)\rho(p_2)
		\left(
			\mP_1^{\mu\nu*}-f_1^{\mu\nu*}
		\right)
		I_\mnrs
		\left(
			\mP_2^{\rho\sigma}-f_2^{\rho\sigma}
		\right)
	\right\rbrace
\\ &\quad\times
	\exp\left\lbrace
		\frac{\kappa}{2}
		\int\td{p}\td{k}\rho(p)
		\left(
			\mP^{\mu\nu*}-f^{\mu\nu*}
		\right)
		a^\dagger_{\mu\nu}
	\right\rbrace
\\ &\quad\times
	\exp\left\lbrace
		-\frac{\kappa}{2}
		\int\td{p}\td{k}\rho(p)
		\left(
			\mP^{\mu\nu}-f^{\mu\nu}
		\right)
		a_{\mu\nu}
	\right\rbrace ,
\end{split}
\end{align}
where $\mP^{\mu\nu}_i\equiv \mP^{\mu\nu}(p_i,k)$ and $f^{\mu\nu}_i\equiv f^{\mu\nu}(p_i,k)$ for $i = 1,2$.
The first exponential involves an integrand of the form
\begin{align}
\label{cIc}
\frac{1}{\wk}
\left(
	\mP_1^{\mu\nu*}-f_1^{\mu\nu*}
\right)
I_\mnrs
\left(
	\mP_2^{\rho\sigma}-f_2^{\rho\sigma}
\right)
\quad\xrightarrow{k\rightarrow 0}\quad
\frac{1}{\wk^3}
c^{\mu\nu*}(p_1,k)
I_\mnrs
c^{\rho\sigma}(p_2,k).
\end{align}
Due to the gauge constraint \eqref{eqn:c1}, the leading term of $c_\mn$ cannot cancel any poles $1/\wk$ in this limit, 
meaning that the integral exhibits IR divergence unless the leading term of \eqref{cIc} in $k$ vanishes.
Using \eqref{c_real} to write $c^{\mn*}= c^\mn+O(k)$ in \eqref{cIc},
we find that the following constraint,
\begin{align}
\label{ccond2}
c^{\mu\nu}(p_1,k)
I_\mnrs
c^{\rho\sigma}(p_2,k) = O(k)
\quad\text{for all $p_1$ and $p_2$,}
\end{align}
is sufficient for $e^{R(t)-R_f}$ to form a unitary operator.
Notice that when this is satisfied, the last two exponentials \eqref{eqn:e_diff} form unitary operators in the Fock space as well.

The subleading $O(k)$ terms in \eqref{ccond2}, which also include the subleading terms in \eqref{c_real} due to rewriting
	$c^{\mn*}= c^\mn+O(k)$, give a finite value to the $k$-integral in \eqref{eqn:e_diff}.
These terms therefore only contribute to the
	normalization of the states and can be ignored.

\section{Cancellation of infrared divergence}
\label{app_ir}
By constructing asymptotic states analogous to that of Faddeev and Kulish \cite{Kulish:1970ut}, 
IR divergence in gravity was shown \cite{Ware:2013zja} to cancel to all loop orders for single-particle asymptotic states
	by making a convenient choice of $c_{\mu\nu}$, i.e. $c^{\mu\nu}(p,k)\epsilon^\pm_{\mu\nu}(k)=0$.
Here we generalize this to multi-particle asymptotic states using a general $c_{\mu\nu}$ that is only subject to
	the basic constraints \eqref{eqn:c1}-\eqref{eqn:cc3}.
We will set $\phi(p,k)=1$ without any loss of generality since this only changes the overall normalization of the states.

The equations involved will turn out to be cumbersome, so let us begin by laying down some shorthand notations.
We remind the reader that the dressed creation and annihilation operators of the scalar particle take the form
\begin{align}
e^{R_f(p)}b^\dagger(p)&=\exp\left[\frac{\kappa}{2}\int\td{k}
\left(f_{\mu\nu}(p,k)a^{\dagger\mu\nu}(k)-f_{\mu\nu}(p,k)a^{\mu\nu}(k)\right)
\right]b^\dagger(p)\\
e^{-R_f(p)}b(p)&=\exp\left[-\frac{\kappa}{2}\int\td{k}
\left(f_{\mu\nu}(p,k)a^{\dagger\mu\nu}(k)-f_{\mu\nu}(p,k)a^{\mu\nu}(k)\right)
\right]b(p)\ .
\end{align}
The dressings $e^{\pm R_f(p)}$ commute with the undressed operators $b$, $b^\dagger$.
If we define
\begin{align}
S_{\mu\nu}(p,k) &= \frac{\kappa}{2}f_{\mu\nu}(p,k)\\
P_{\mu\nu}(p,k) &= \frac{\kappa}{2}\left(\frac{p_\mu p_\nu}{p \cdot k}\right)\\
C_{\mu\nu}(p,k) &= \frac{\kappa}{2}\frac{c_{\mu\nu}(p,k)}{\wk}
\end{align}
so that $S_{\mu\nu}=P_{\mu\nu}+C_{\mu\nu}$, we have
\begin{align}
e^{R_f(p)} b^\dagger(p)
&=\exp\left[\int\td{k}
\left(S_{\mu\nu}(p,k)a^{\dagger\mu\nu}(k)-S_{\mu\nu}(p,k)a^{\mu\nu}(k)\right)
\right]b^\dagger(p)\\
e^{-R_f(p)} b(p)
&=\exp\left[-\int\td{k}
\left(S_{\mu\nu}(p,k)a^{\dagger\mu\nu}(k)-S_{\mu\nu}(p,k)a^{\mu\nu}(k)\right)
\right]b(p).
\end{align}
We will use the superscript ``in" (``out") to denote the quantity summed over all incoming (outgoing)
	scalar particles. The superscript ``tot" will denote the difference between ``out" and ``in". For example,
\begin{align}
S^\text{in}_{\mu\nu}(k) &= \sum_{i\in\text{in}} S_{\mu\nu}(p_i,k), \quad
S^\text{out}_{\mu\nu}(k) = \sum_{i\in\text{out}} S_{\mu\nu}(p_i,k), \quad
S^\text{tot}_{\mu\nu} = S^\text{out}_{\mu\nu} - S^\text{in}_{\mu\nu}, \\
P^\text{in}_{\mu\nu}(k) &= \sum_{i\in\text{in}} P_{\mu\nu}(p_i,k), \quad
P^\text{out}_{\mu\nu}(k) = \sum_{i\in\text{out}} P_{\mu\nu}(p_i,k), \quad
P^\text{tot}_{\mu\nu} = P^\text{out}_{\mu\nu} - P^\text{in}_{\mu\nu}, \\
C^\text{in}_{\mu\nu}(k) &= \sum_{i\in\text{in}} C_{\mu\nu}(p_i,k), \quad
C^\text{out}_{\mu\nu}(k) = \sum_{i\in\text{out}} C_{\mu\nu}(p_i,k), \quad
C^\text{tot}_{\mu\nu} = C^\text{out}_{\mu\nu} - C^\text{in}_{\mu\nu}.
\end{align}
We will sometimes write
\begin{align}
S^n_{\mu\nu} &\equiv S_{\mu\nu}(p_n,k)
\end{align}
in contexts where the graviton momentum $k$ is unambiguous.

\subsection{Sources of infrared divergence}
Listed below are the possible sources of IR divergence:
\begin{enumerate}
	\item Virtual gravitons.
		It is well known that only the virtual gravitons connecting two external legs
			produce IR divergence, and that their contribution exponentiates \cite{Weinberg:1965nx}.
			This contribution takes the form \cite{Ware:2013zja}
		\begin{align}
		\label{eqn:virt_contr}
		\exp\left[-\frac{\kappa^2}{128\pi^3}\sum_{n,m}\int\frac{ d^3k}{\wk}
			\frac{\eta_n\eta_m\left[(p_n\cdot p_m)^2-(1/2)p_n^2p_m^2\right]}{(p_n\cdot k)(p_m\cdot k)}\right],
		\end{align}
		where each sum runs over the external particles. $\eta=+1$ for an outgoing particle, and $\eta=-1$ for an incoming particle.
	\item Real gravitons. External soft gravitons are another source of IR divergence \cite{Weinberg:1965nx}.
		In this section the external states will involve gravitons only in the form of Faddeev-Kulish clouds.
	\item Interacting gravitons. We reserve the term ``interacting" to denote the gravitons that connect a Faddeev-Kulish
		cloud to either an external or an internal leg.
		We follow the procedure analogous to the work of Chung \cite{Chung:1965zza} to factor out the IR divergence
		from this type of contribution.
	\item Cloud-to-cloud gravitons.
		These gravitons propagate from one cloud to another. We can further group these into two types:
	\begin{enumerate}
		\item ``Disconnected" gravitons. We will use this term to denote gravitons that connect
			the cloud of an incoming particle with the cloud of an outgoing particle.
		\item In-to-in/out-to-out gravitons. In-to-in (out-to-out) gravitons connect two incoming (outgoing) clouds.
			Note that the graviton can be emitted and absorbed by the same cloud,
				see figures \ref{fig:1ptl_cloud}(b) and \ref{fig:1ptl_cloud}(c).
	\end{enumerate}
\end{enumerate}

\subsection{Single-particle external states, cancellation to one loop}
We start with the case of single-scalar in, single-scalar out,
and show that the divergent factors cancel to second-order in the interaction.
In the next subsection we will see how this generalizes to multiple-scalar in, multiple-scalar out, and show
the cancellation to all orders of interaction.

Consider the single-scalar asymptotic in-state
\begin{align}
\ket{\text{i}} &= e^{R_f(p_i)} b^\dagger(p_i) \ket{0}\\
&=\exp\left[\int
	\td{k}
	\left(S^i_{\mu\nu}a^{\dagger\mu\nu}-S^i_{\mu\nu}a^{\mu\nu}\right)\right]
b^\dagger(p_i) \ket{0}.
\end{align}
The commutator
\begin{align}
&\left[
	\left(\int\td{k}
		S^i_{\mu\nu}a^{\dagger\mu\nu}\right),
	\left(-\int\td{k}
		S^i_{\mu\nu}a^{\mu\nu}\right)
\right]
=\frac{1}{2}\int \td{k}S^i_{\mu\nu}I^{\mnrs}S^i_{\rho\sigma}
\end{align}
is a c-number, so we can use the BCH formula $e^{A+B}=e^Ae^Be^{-\frac{1}{2}[A,B]}$ to write
\begin{align}
\begin{split}
&\exp\left[
	\int\td{k}
	\left(S^i_{\mu\nu}a^{\dagger\mu\nu}-S^i_{\mu\nu}a^{\mu\nu}\right)
\right]\\
&=\exp\left(\int\td{k}
	S^i_{\mu\nu}a^{\dagger\mu\nu}\right)
	\exp\left(-\int\td{k}
	S^i_{\mu\nu}a^{\mu\nu}\right)
	\exp\left(
		-\frac{1}{4}\int \td{k}S^i_{\mu\nu}I^{\mnrs}S^i_{\rho\sigma}
	\right).
\end{split}
\end{align}
Therefore, the in-state may be written as
\begin{align}
\ket{\text{i}} &=
\exp\left(-\frac{1}{4}\int \td{k}
	S^i_{\mu\nu} I^{\mnrs} S^i_{\rho\sigma}
\right)
\exp\left(\int\td{k}
	S^i_{\mu\nu}a^{\dagger\mu\nu}\right)
	b^\dagger(p_i)\ket{0},
\end{align}
since $a^{\mu\nu}$ commutes with $b^\dagger$ and annihilates the vacuum. To the lowest order, this is
\begin{align}
\label{eqn:1ptl_in}
\ket{\text{i}} =
\left(
	1
	-\frac{1}{4}\int \td{k}
		S^{i}_{\mu\nu}I^{\mnrs}S^{i}_{\rho\sigma}
	+\int\td{k}
		S^{i}_{\mu\nu}a^{\dagger\mu\nu}
\right)
b^\dagger(p_i) \ket{0}.
\end{align}
Similarly, we may write the asymptotic out-state as
\begin{align}
\label{eqn:1ptl_out}
\bra{\text{f}} &= \bra{0} b(p_f) e^{-R_f(p_f)}\\
&= \bra{0} b(p_f)
	\exp\left[-\int\td{k}
		\left(S^{f}_{\mu\nu}a^{\dagger\mu\nu}-S^f_{\mu\nu}a^{\mu\nu}\right)\right]\\
&=\bra{0} b(p_f)
\exp\left(\int\td{k}
	S^{f}_{\mu\nu}a^{\mu\nu}\right)
\exp\left(-\frac{1}{4}\int \td{k}S^f_{\mu\nu}I^{\mnrs}S^{f}_{\rho\sigma}\right),
\end{align}
or to the lowest order,
\begin{align}
\bra{\text{f}} = \bra{0}b(p_f)\left(1
-\frac{1}{4}\int \td{k}S^{f}_{\mu\nu}I^{\mnrs}S^{f}_{\rho\sigma}
+\int\td{k}S^{f}_{\mu\nu}a^{\mu\nu}
\right).
\end{align}
We will now demonstrate that an amplitude of the form $\braket{\text{f}|\mS|\text{i}}$ is free of IR divergence.

\begin{figure}[t]
\centering
    \begin{subfigure}[b]{0.23\textwidth}
		\includegraphics[width=\textwidth]{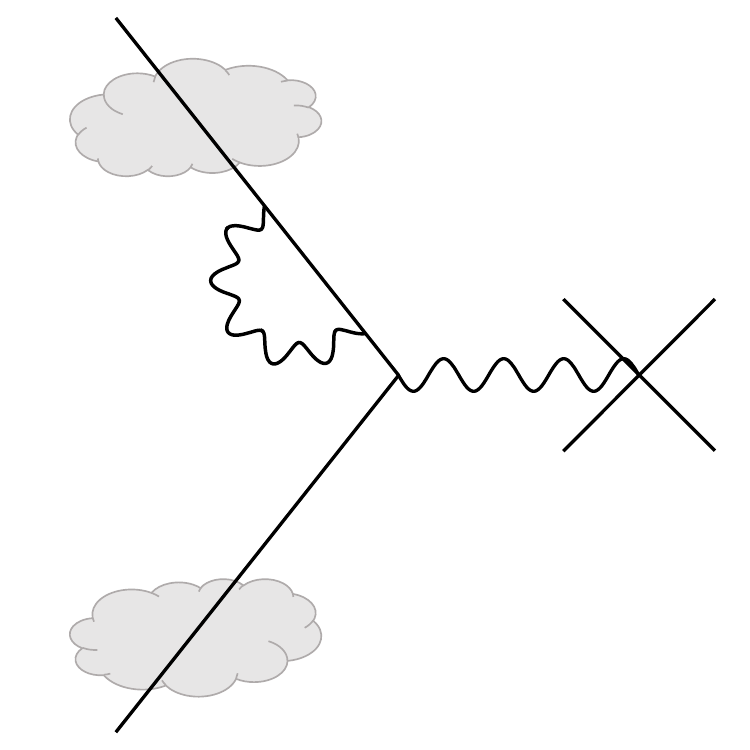}
		\caption{}
    \end{subfigure}
    \begin{subfigure}[b]{0.23\textwidth}
		\includegraphics[width=\textwidth]{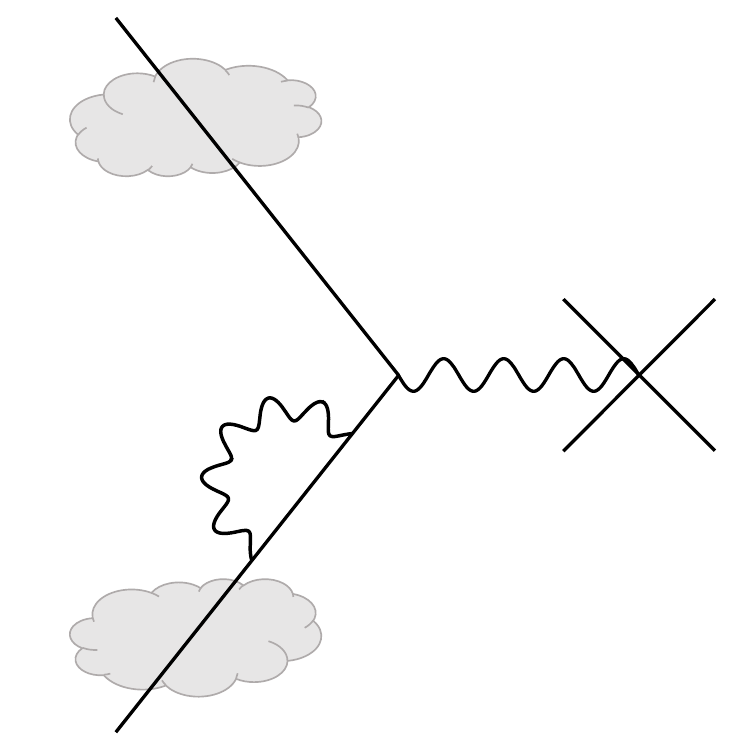}
		\caption{}
    \end{subfigure}
    \begin{subfigure}[b]{0.23\textwidth}
		\includegraphics[width=\textwidth]{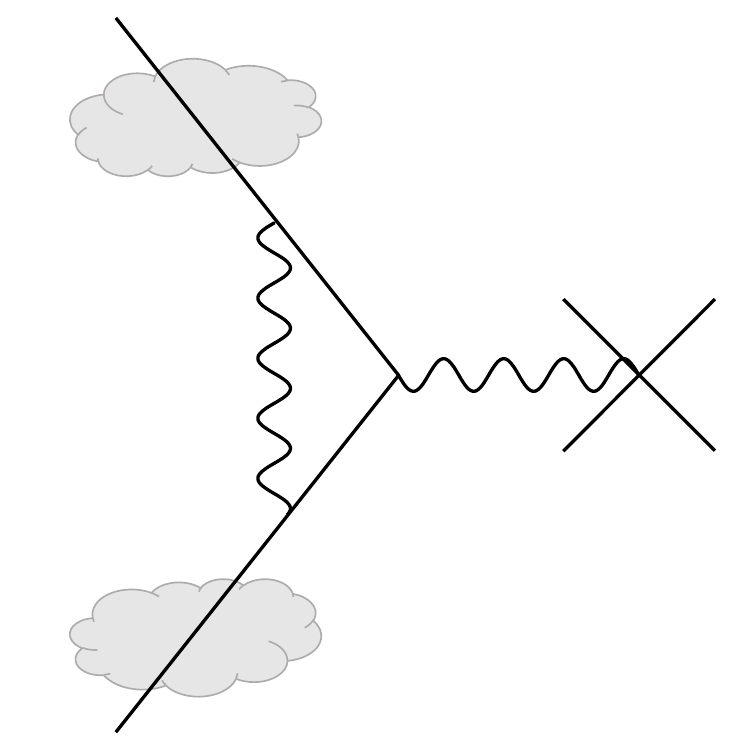}
		\caption{}
    \end{subfigure}
\caption{Contributions of a virtual graviton.}
\label{fig:1ptl_virt}
\end{figure}

Let us begin with the contribution of virtual gravitons.
Diagrams that fall into this category are given in figure \ref{fig:1ptl_virt}.
\eqref{eqn:virt_contr} sums up these contributions,
	which in our case of single particle external states can be written as
\begin{align}
&\exp\left[-\kappa^2\sum_{n,m}\int\frac{ d^3k}{128\pi^3\wk}
	\frac{\eta_n\eta_m\left[(p_n\cdot p_m)^2-(1/2)p_n^2p_m^2\right]}{(p_n\cdot k)(p_m\cdot k)}\right]\\
&\approx 1-\kappa^2\sum_{n,m}\int\frac{ d^3k}{128\pi^3\wk}
\frac{\eta_n\eta_m\left[(p_n\cdot p_m)^2-(1/2)p_n^2p_m^2\right]}{(p_n\cdot k)(p_m\cdot k)} \\
&= 1+\frac{\kappa^2}{128\pi^3}\int\frac{ d^3k}{\wk}\left[
\frac{p^4_f}{2(p_f\cdot k)^2}+\frac{p^4_i}{2(p_i\cdot k)^2}
-2\left(\frac{(p_f\cdot p_i)^2-\frac{1}{2}p_f^2p_i^2}{(p_f\cdot k)(p_i\cdot k)}\right)
\right].
\end{align}
Thus we find the contribution $\Afactor^{(1)}_\text{virt}$ of virtual gravitons to be
\begin{align}
\Afactor_\text{virt}^{(1)}=
\frac{\kappa^2}{128\pi^3}\int\frac{ d^3k}{\wk}\left[
\frac{p^4_f}{2(p_f\cdot k)^2}+\frac{p^4_i}{2(p_i\cdot k)^2}
-2\left(\frac{(p_f\cdot p_i)^2-\frac{1}{2}p_f^2p_i^2}{(p_f\cdot k)(p_i\cdot k)}\right)
\right],
\end{align}
where the superscript $(1)$ emphasizes that this is the leading term in the interaction.

\begin{figure}[t]
	\centering
    \begin{subfigure}[b]{0.23\textwidth}
		\includegraphics[width=\textwidth]{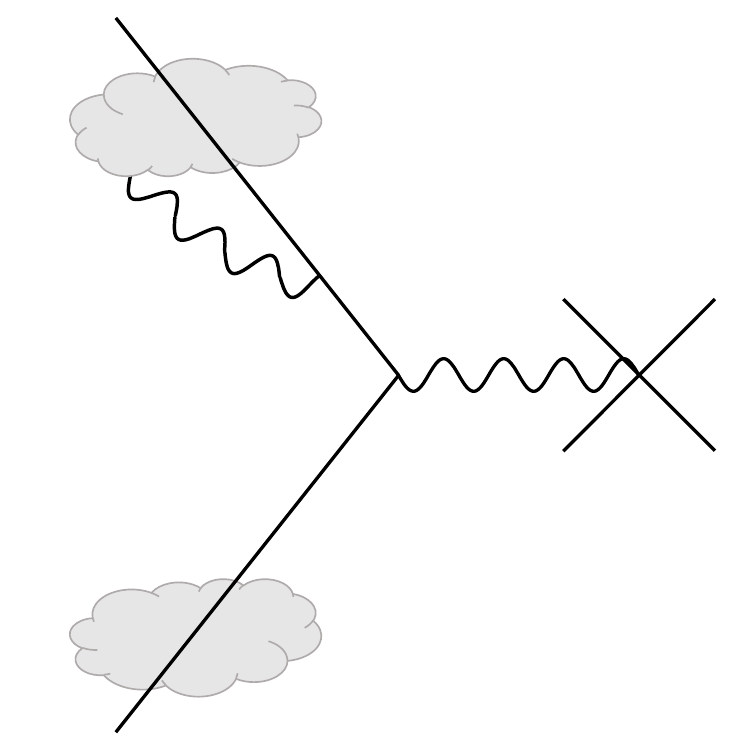}
		\caption{}
    \end{subfigure}
    \begin{subfigure}[b]{0.23\textwidth}
		\includegraphics[width=\textwidth]{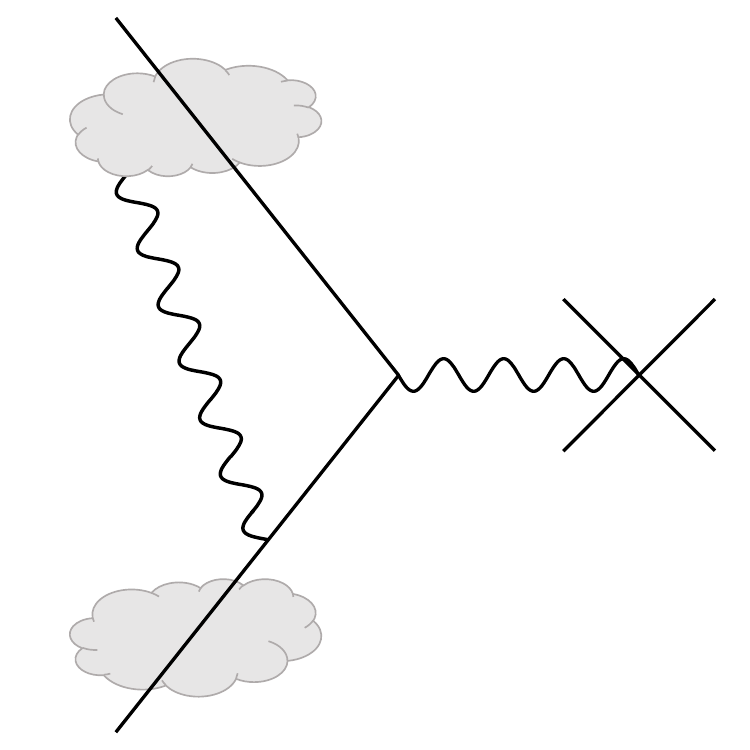}
		\caption{}
    \end{subfigure}
    \begin{subfigure}[b]{0.23\textwidth}
		\includegraphics[width=\textwidth]{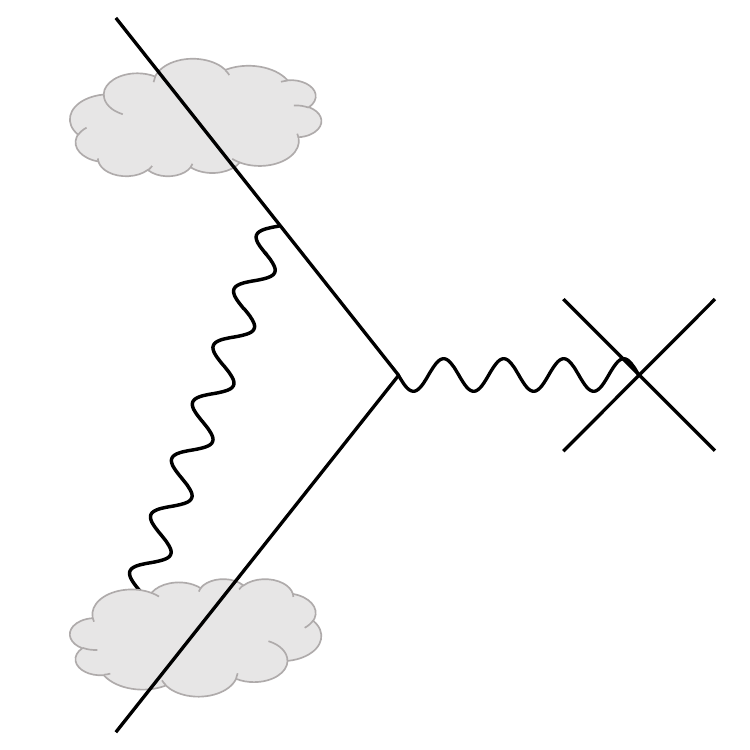}
		\caption{}
    \end{subfigure}
    \begin{subfigure}[b]{0.23\textwidth}
		\includegraphics[width=\textwidth]{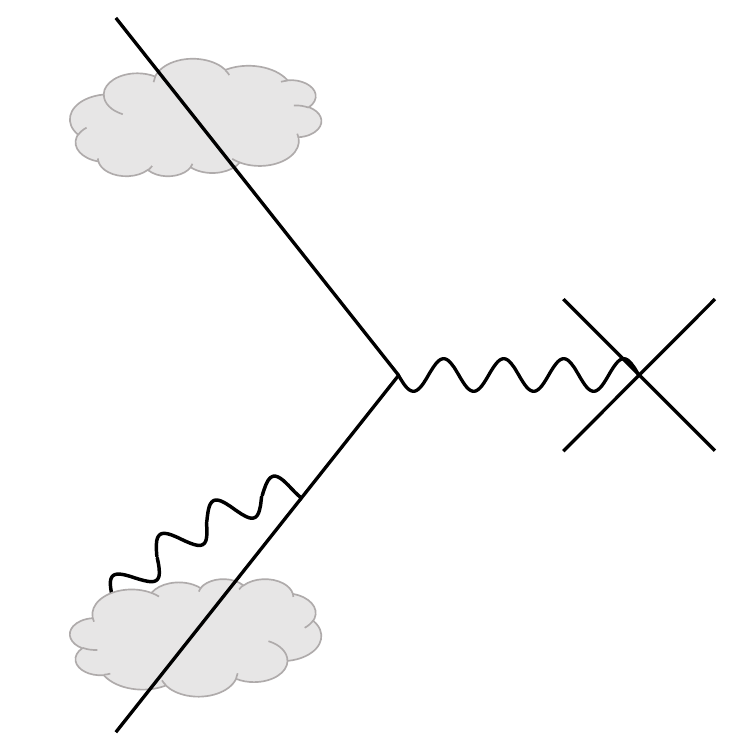}
		\caption{}
    \end{subfigure}
	\caption{Contributions of interacting gravitons.}
	\label{fig:1ptl_int}
\end{figure}

Next, we consider the contributions of interacting gravitons.
There are four diagrams that are IR-divergent, which are shown in figure \ref{fig:1ptl_int}.
Contribution from figure \ref{fig:1ptl_int}(a) yields a factor of
\begin{align}
\label{eqn:int_1}
\int\td{k}
	S^f_{\mu\nu}
	\frac{1}{2}I^{\mnrs}
	\left(\frac{-i}{2p^f\cdot k}\right)\left(i\kappa p^f_{\rho} p^f_{\sigma}\right)
	=\frac{1}{2}\int \td{k} S^f_{\mu\nu}I^{\mnrs}P^f_{\rho\sigma}\ ,
\end{align}
where $-i/2p^f\cdot k$ is the propagator,
$i\kappa p^f_{\rho} p^f_{\sigma}$ comes from the vertex rule,
and the rest comes from the contraction of an
outgoing cloud and $h^{\rho\sigma}(x)$. Similarly, diagrams (b), (c) and (d) contribute the following factors respectively:
\begin{align}
-&\frac{1}{2}\int \td{k} S^f_{\mu\nu}I^{\mnrs}P^i_{\rho\sigma}\\
-&\frac{1}{2}\int \td{k} S^i_{\mu\nu}I^{\mnrs}P^f_{\rho\sigma}\\
\label{eqn:int_4}
&\frac{1}{2}\int \td{k} S^i_{\mu\nu}I^{\mnrs}P^i_{\rho\sigma}
\end{align}
The net contribution of interacting graviton is the sum of \eqref{eqn:int_1}-\eqref{eqn:int_4}, which reads
\begin{align}
\label{eqn:secondhalf}
\frac{1}{2}\int \td{k} \left(S^f_{\mu\nu}-S^i_{\mu\nu}\right)I^{\mnrs}\left(P^f_{\rho\sigma}-P^i_{\rho\sigma}\right)
	= \frac{1}{2}\int \td{k}S^\text{tot}_{\mu\nu}I^{\mnrs}P^\text{tot}_{\rho\sigma}\ .
\end{align}

\begin{figure}[t]
	\centering
    \begin{subfigure}[b]{0.23\textwidth}
		\includegraphics[width=\textwidth]{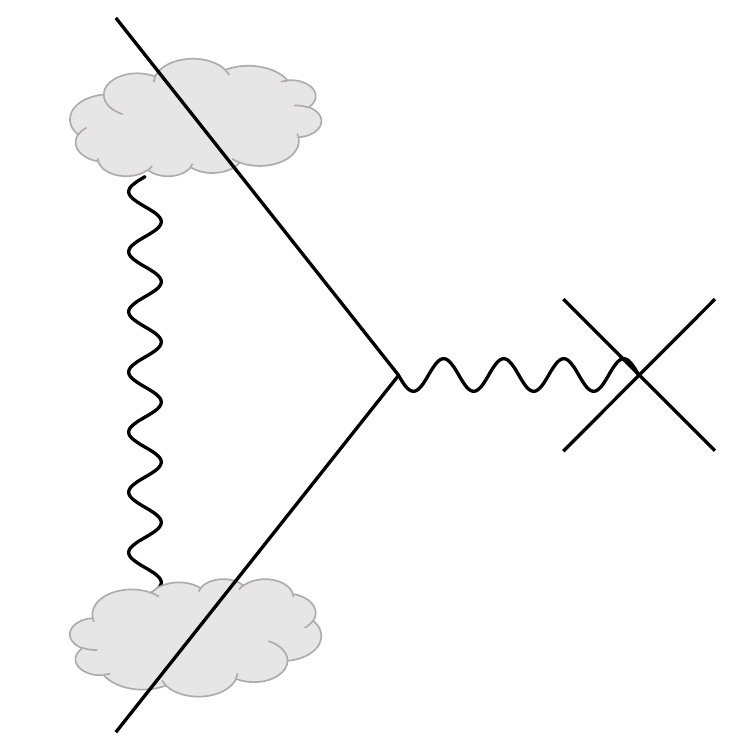}
		\caption{}
    \end{subfigure}
    \begin{subfigure}[b]{0.23\textwidth}
		\includegraphics[width=\textwidth]{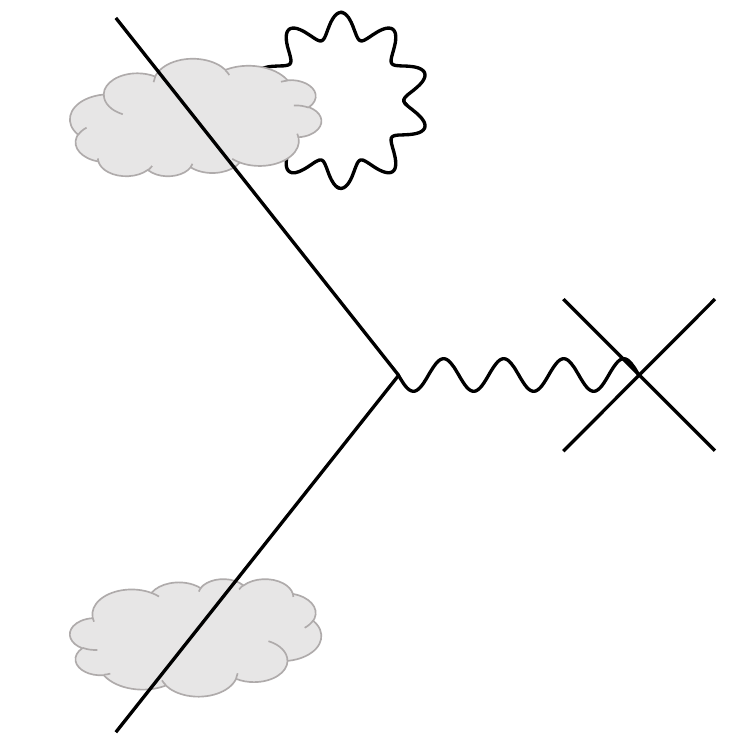}
		\caption{}
    \end{subfigure}
    \begin{subfigure}[b]{0.23\textwidth}
		\includegraphics[width=\textwidth]{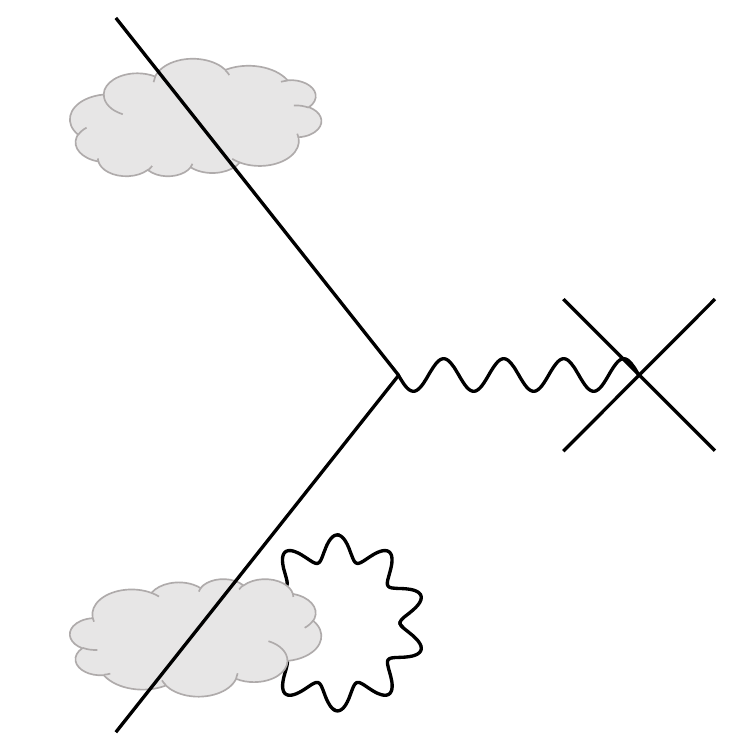}
		\caption{}
    \end{subfigure}
	\caption{Contributions of cloud-to-cloud gravitons.}
	\label{fig:1ptl_cloud}
\end{figure}

The last contribution comes from the cloud-to-cloud gravitons.
There are three diagrams that correspond to this category, shown in figure \ref{fig:1ptl_cloud}. 
Figure \ref{fig:1ptl_cloud}(a) shows the ``disconnected" graviton line.
Recalling that the initial and final states are
\begin{align}
\label{eqn:in_repeated}
\ket{\text{i}} &=
\left(
1
-\frac{1}{4}\int \td{k}S^{i}_{\mu\nu}I^{\mnrs}S^{i}_{\rho\sigma}
+\int\td{k}S^{i}_{\mu\nu}a^{\dagger\mu\nu}
\right)
b^\dagger(p_i) \ket{0}\\
\label{eqn:out_repeated}
\bra{\text{f}} &= \bra{0}b(p_f)\left(1
-\frac{1}{4}\int \td{k}S^{f}_{\mu\nu}I^{\mnrs}S^{f}_{\rho\sigma}
+\int\td{k}S^{f}_{\mu\nu}a^{\mu\nu}
\right),
\end{align}
we can see that the disconnected line corresponds to the contraction
	of the last terms of \eqref{eqn:in_repeated} and \eqref{eqn:out_repeated},
\begin{align}
&\int\td{k}\td{k'}
	S_{\mu\nu}(p_f,k)S_{\rho\sigma}(p_i,k')\braket{0|a^{\mu\nu}(k)a^{\dagger\rho\sigma}(k')|0}\\
&=\frac{1}{2}\int \td{k}S^f_{\mu\nu}I^{\mnrs}S^{i}_{\rho\sigma}\\
\label{eqn:disccontrib}
&=\frac{1}{4}\int \td{k}S^f_{\mu\nu}I^{\mnrs}S^{i}_{\rho\sigma}
+\frac{1}{4}\int \td{k}S^i_{\mu\nu}I^{\mnrs}S^{f}_{\rho\sigma}\ ,
\end{align}
where in the last equation we used the symmetry of
$I^{\mnrs}=\eta^{\mu\rho}\eta^{\nu\sigma}+\eta^{\mu\sigma}\eta^{\nu\rho}
-\eta^{\mu\nu}\eta^{\rho\sigma}$ under $(\mu\nu)\leftrightarrow(\rho\sigma)$.
Figures \ref{fig:1ptl_cloud}(b) and \ref{fig:1ptl_cloud}(c) are the out-to-out and in-to-in graviton lines, respectively. These 
contribute a factor coming from the second terms of \eqref{eqn:out_repeated} and \eqref{eqn:in_repeated},
\begin{align}
\label{eqn:quadcontrib}
-\frac{1}{4}\int \td{k}\left(
S^{f}_{\mu\nu}I^{\mnrs}S^{f}_{\rho\sigma}+S^{i}_{\mu\nu}I^{\mnrs}S^{i}_{\rho\sigma}
\right),
\end{align}
which, combined with \eqref{eqn:disccontrib}, form the cloud-to-cloud contribution
\begin{align}
\label{eqn:firsthalf}
-\frac{1}{4}\int \td{k}\left(S^f_{\mu\nu}-S^i_{\mu\nu}\right) I^{\mnrs} \left(S^f_{\rho\sigma}-S^i_{\rho\sigma}\right)
= -\frac{1}{4}\int \td{k}S^\text{tot}_{\mu\nu} I^{\mnrs} S^\text{tot}_{\rho\sigma}\ .
\end{align}

The leading contribution $\Afactor_\text{cloud}^{(1)}$ involving the clouds can therefore be written as the sum of
	\eqref{eqn:secondhalf} and \eqref{eqn:firsthalf}:
\begin{align}
\label{eqn:totalcloud}
\Afactor_\text{cloud}^{(1)}=-\frac{1}{4}\int \td{k} S^\text{tot}_{\mu\nu} I^{\mnrs} S^\text{tot}_{\rho\sigma}
	+\frac{1}{2}\int \td{k} S^\text{tot}_{\mu\nu} I^{\mnrs} P^\text{tot}_{\rho\sigma}\ .
\end{align}
Noting that $S^\text{tot}_{\mu\nu}=P^\text{tot}_{\mu\nu}+C^\text{tot}_{\mu\nu}$, we write
\begin{align}
\Afactor_\text{cloud}^{(1)}
&=\frac{1}{4}\int \td{k} 
I^\mnrs
\Big[
-\left(P^\text{tot}_\mn+C^\text{tot}_\mn\right)
\left(P^\text{tot}_{\rho\sigma}+C^\text{tot}_{\rho\sigma}\right)
+2\left(P^\text{tot}_\mn+C^\text{tot}_\mn\right)P^\text{tot}_{\rho\sigma}
\Big]
\\
&=
\frac{1}{4}\int \td{k} P^\text{tot}_{\mu\nu}I^{\mnrs}P^\text{tot}_{\rho\sigma}
-\frac{1}{4}\int \td{k} C^\text{tot}_{\mu\nu}I^{\mnrs}C^\text{tot}_{\rho\sigma}.
\end{align}
The second term involving the integral
\begin{align}
\label{1ptl_cIc}
\int\frac{d^3k}{\wk^3}
\left(c^f_\mn-c^i_\mn\right)
I^\mnrs
\left(c^f_\rs-c^i_\rs\right),
\end{align}
derives solely from the interactions between graviton clouds.
Note that in this case of single-particle states, we cannot use different $c_\mn$ for the incoming and outgoing particles,
	since that will render the integral \eqref{1ptl_cIc} divergent.
This point will become more clear when we study the case of multi-particle states in the next subsection.
This term thus vanishes due to the convergence constraint \eqref{eqn:cc3}.
Then we are left with
\begin{align}
\Afactor_\text{cloud}^{(1)}
&=
\frac{1}{4}\int \td{k} P^\text{tot}_{\mu\nu}I^{\mnrs}P^\text{tot}_{\rho\sigma}\\
&=\frac{\kappa^2}{16}
\int\frac{ d^3k}{(2\pi)^3 2\wk}
\left(
	\frac{p^f_\mu p^f_\nu}{p^f\cdot k}
	-\frac{p^i_\mu p^i_\nu}{p^i\cdot k}
\right)
I^{\mnrs}
\left(
	\frac{p^f_\rho p^f_\sigma}{p^f\cdot k}
	-\frac{p^i_\rho p^i_\sigma}{p^i\cdot k}
\right)
\\
&=\frac{\kappa^2}{128\pi^3}\int\frac{ d^3k}{\wk}\left[
\frac{p^4_f}{2(p_f\cdot k)^2}+\frac{p^4_i}{2(p_i\cdot k)^2}
-2\left(\frac{(p_f\cdot p_i)^2-\frac{1}{2}p_f^2p_i^2}{(p_f\cdot k)(p_i\cdot k)}\right)
\right].
\end{align}
This is precisely $\Afactor_\text{virt}^{(1)}$ with the opposite sign,
and therefore cancels the contribution of the virtual gravitons.

\subsection{Multi-particle external states, cancellation to all orders}
\noindent
To all loop orders, the contribution $\Afactor_\text{virt}$ of soft gravitons in loops is given by \eqref{eqn:virt_contr}, which reads
\begin{align}
\label{eqn:Avirt}
\Afactor_\text{virt}
&=\exp\left[-\frac{\kappa^2}{128\pi^3}\sum_{n,m}\int\frac{ d^3k}{\wk}
\frac{\eta_n\eta_m\left[(p_n\cdot p_m)^2-(1/2)p_n^2p_m^2\right]}{(p_n\cdot k)(p_m\cdot k)}\right]\\
&=\exp\left(-\frac{1}{4}\sum_{n,m}\eta_n\eta_m\int \td{k}P^n_{\mu\nu}I^{\mnrs}P^m_{\rho\sigma}\right),
\end{align}
where the summation indices $n$ and $m$ run over all external particles.

Next we compute the interacting gravitons' contribution.
To this end, let us first examine how the insertion of a soft graviton affects the amplitude of a diagram,
following the procedure analogous to that of Chung \cite{Chung:1965zza} for QED.
Suppose we have a diagram with amplitude $M^{(0)}$ that does not contain any soft gravitons.
Inserting a soft graviton $a_{\mu_1\nu_1}(k_1)$ or $a^\dagger_{\mu_1\nu_1}(k_1)$ will give us a new amplitude
\begin{align}
M^{(1)}_{\mu_1\nu_1}(k_1) = \pm P^\text{tot}_{\mu_1\nu_1}(k)M^{(0)} + \tilde{\xi}_{\mu_1\nu_1}(k_1),
\end{align}
where the net soft factor $P^\text{tot}_{\mu_1\nu_1}(k)$ comes from attaching the graviton to the external legs,
and $\tilde{\xi}_{\mu_1\nu_1}(k_1)$ comes from attaching it to
the body of the diagram and does not contain IR divergence in $k_1$. The $+$ ($-$) sign corresponds to emission (absorption)
of the graviton. The Lorentz indices $\mu_1$ and $\nu_1$ will eventually contract with the clouds
$\int \td{k}S_{\rho\sigma}\frac{1}{2}I^{\rsmn}$, but we will leave them free for now.
We can see that an amplitude $M^{(n)}_{\mu_1\nu_1\cdots\mu_n\nu_n}$ with $n$ real soft gravitons may be written as
\begin{align}
\begin{split}
M^{(n)}_{\mu_1\nu_1\cdots\mu_n\nu_n}(k_1,\cdots,k_n)
	= \pm P^\text{tot}_{\mu_n\nu_n}(k_n)
	&M^{(n-1)}_{\mu_1\nu_1\cdots\mu_{n-1}\nu_{n-1}}(k_1,\cdots,k_{n-1})
	\\&\qquad
	+ \tilde{\xi}_{\mu_1\nu_1\cdots\mu_n\nu_n}(k_1,\cdots,k_{n-1};k_n),
\end{split}
\end{align}
where $\tilde{\xi}_{\mu_1\nu_1\cdots\mu_n\nu_n}(k_1,\cdots,k_{n-1};k_n)$ does not contain IR divergence in $k_n$.
We know from \cite{YENNIE1961379} that such equation can be unwound as a sum over all permutations of
the gravitons, in this case represented by the labels ($\mu,\nu,k$)'s:
\begin{align}
M^{(n)}_{\mu_1\nu_1\cdots\mu_n\nu_n}(k_1,\cdots,k_n)
	&= 
	\sum_{s=0}^n
	\sum_{\substack{\text{Perm}\\(\mu,\nu,k)}}
	\frac{(-1)^m}{s!(n-s)!}
	\left[\prod_{i=1}^sP^\text{tot}_{\mu_i\nu_i}(k_i)\right]
	\xi_{\mu_{s+1}\nu_{s+1}\cdots\mu_n\nu_n}(k_{s+1},\cdots,k_n),
\end{align}
where $m$ is the number of absorbed gravitons and
	$\xi$'s are some IR-convergent functions symmetric in the gravitons, or equivalently in the labels $(\mu,\nu,k)$'s.

\begin{figure}[t]
	\centering
	\includegraphics[width=.45\textwidth]{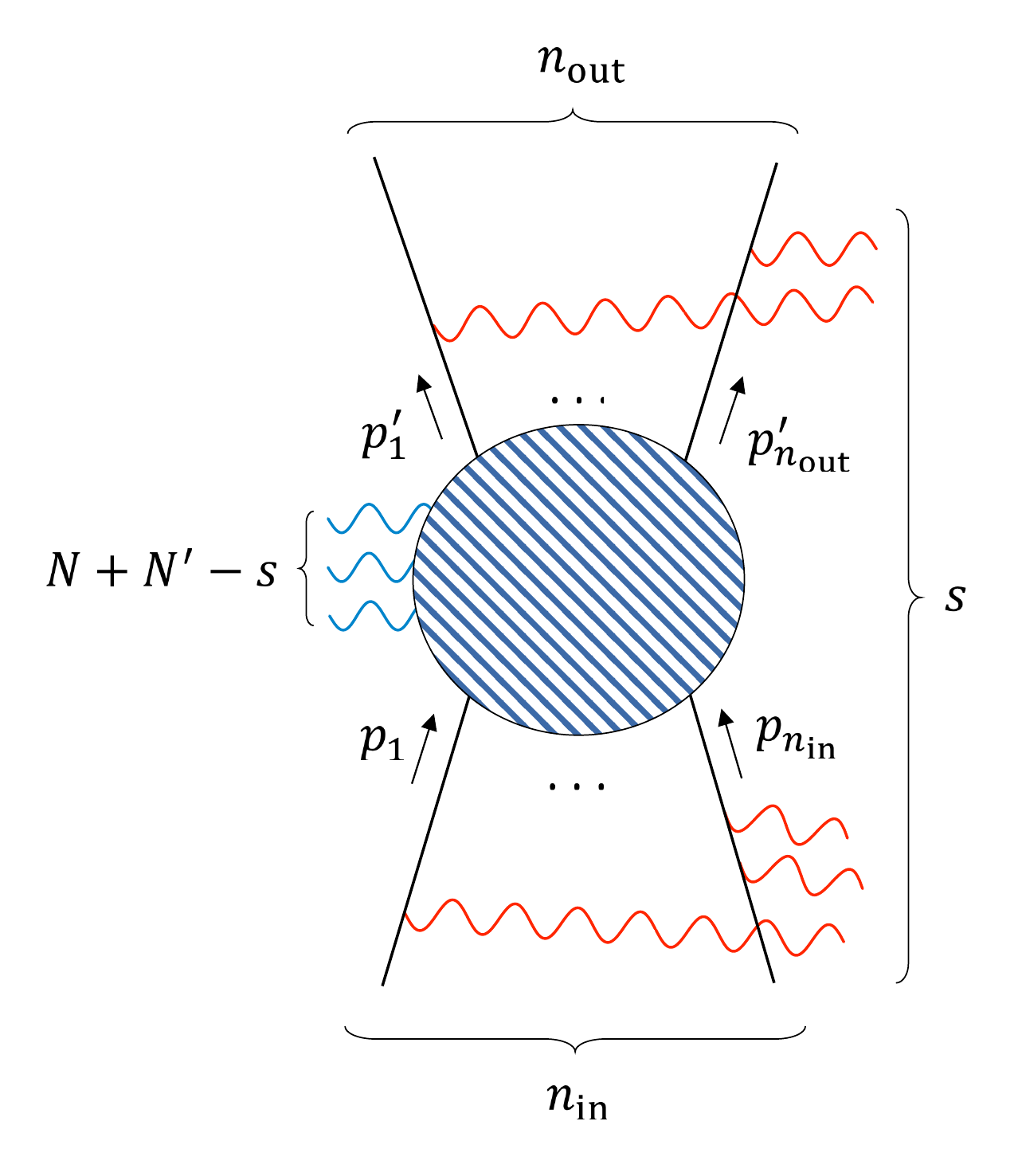}
	\caption{A diagram with $n_\text{in}$ incoming, $n_\text{out}$ outgoing scalar particles, and $N+N'$ interacting gravitons.
	The loose ends of graviton lines connect to the clouds, which are not drawn here.
	There are $s$ gravitons (colored red) connected to external legs,
		each contributing an IR-divergent factor $\pm P_{\mu\nu}$.
	The remaining $N+N'-s$ gravitons (colored blue) connect to the internal legs and constitute the
		IR-convergent part $\xi$.
	}
	\label{fig:mult_body}
\end{figure}

We will examine the amplitude of a diagram with $N$ ($N'$) interacting soft gravitons that connect to
	the clouds of incoming (outgoing) scalars.
This puts $n=N+N'$, so let us write
\begin{align}
M^{(N+N')}_{\mu_1\nu_1\cdots\mu_{N+N'}\nu_{N+N'}}(k_1,\cdots,k_{N+N'})
	&= 
	(-1)^N
	\sum_{s=0}^{N+N'}
	\sum_{\substack{\text{Perm}\\(\mu,\nu,k)}}
	\frac{M^{(N+N',s)}_{\mu_1\nu_1\cdots\mu_{N+N'}\nu_{N+N'}}(k_1,\cdots,k_{N+N'})}{s!(N+N'-s)!}
\end{align}	
with the restricted amplitude defined by
\begin{align}
M^{(n,s)}_{\mu_1\nu_1\cdots\mu_{n}\nu_{n}}(k_1,\cdots,k_{n})
	&\equiv
	\left[\prod_{i=1}^sP^\text{tot}_{\mu_i\nu_i}(k_i)\right]
	\xi_{\mu_{s+1}\nu_{s+1}\cdots\mu_{n}\nu_{n}}(k_{s+1},\cdots,k_{n}),
\end{align}
representing the sum of all diagrams where
	\textit{the first} $s$ gravitons connect to external legs and the rest to internal legs.
	One such diagram is shown in figure \ref{fig:mult_body}.
	The product $\prod_iP^\text{tot}_{\mu_i\nu_i}$ is the IR-divergent factor due to the gravitons (red in the figure) connecting to external legs.
	The function $\xi_{\mu_{s+1}\nu_{s+1}\cdots}$ is the contribution of the remaining gravitons (blue in the figure) connecting to internal legs.
	One can see that $M^{(n,s)}_{\mu_1\nu_1\cdots}$ is symmetric in the in the first $s$ and the last $N+N'-s$
		labels $(\mu,\nu,k)$. The expression
\begin{align}
	\sum_{\substack{\text{Perm}\\(\mu,\nu,k)}}
	\frac{1}{s!(N+N'-s)!}
	M^{(N+N',s)}_{\mu_1\nu_1\cdots\mu_{N+N'}\nu_{N+N'}}(k_1,\cdots,k_{N+N'})
\end{align}
hence represents the sum of all diagrams that have $N+N'$ interacting gravitons where any $s$ of them
	are connected to the external legs.
Since $M^{(N+N')}_{\mu_1\nu_1\cdots}$ sums over these diagrams for all $0\le s \le N+N'$,
	apart from the factor $(-1)^N$, it represents the amplitude (with loose ends) of a process involving
	$N+N'$ interacting gravitons.

Now we connect the loose ends to the graviton clouds.
Let us restrict our attention to a specific configuration, where the $i$th ($j$th) incoming (outgoing) cloud has $N_i$ ($N'_j$)
	interacting gravitons connected to it, so that $\sum_{i\in\text{in}}N_i=N$ and $\sum_{j\in\text{out}}N'_j=N'$.
Later we will sum over all possible configurations.
As we saw in the case of a single-particle state, connecting a graviton to the cloud of an external particle having
	momentum $p$ amounts to contracting with an expression of the form
\begin{align}
\frac{1}{2}\int \td{k}S_{\mu\nu}(p,k)I^{\mnrs}.
\end{align}
The Lorentz indices $\rho$ and $\sigma$ contract with the indices of the corresponding loose end.
After connecting all loose ends, we will have $N+N'$ copies of these integrals with varying momenta $p$,
	their Lorentz indices contracted with $M^{(N+N')}_{\mu_1\nu_1\cdots}$.
The order in which the gravitons are connected is irrelevant as long as we have the same configuration
	($\lbrace N_i\rbrace$, $\lbrace N'_j\rbrace$), because in $M^{(N+N')}_{\mu_1\nu_1\cdots}$
	we are summing over all permutations of the loose ends.
Since order does not matter, let us simply connect the first $N_1$ gravitons $(\mu_1,\nu_1,k_1),\cdots,(\mu_{N_1},\nu_{N_1},k_{N_1})$
to the first incoming cloud, the next $N_2$ gravitons to the second cloud, and so on.
By the time we exhaust all of the incoming clouds, we would have connected $N$ gravitons, leaving us with $N'$ loose ends.
Then, we repeat this procedure for the outgoing clouds - connect the first $N'_1$ among the $N'$ leftover gravitons to
	the first outgoing cloud, etc.
For notational simplicity, let us define the sequence
\begin{align}
\begin{split}
\label{eqn:seq}
(a_i)_{i=1}^{N+N'}
	&=\bigg(
		\overbrace{
		\underbrace{p_1, \cdots, p_1}_{N_1},
		\underbrace{p_2, \cdots, p_2}_{N_2},
		\cdots,
		\underbrace{p_{n_\text{in}}, \cdots, p_{n_\text{in}}}_{N_{n_\text{in}}},
		}^{N}
		\overbrace{
		\underbrace{p'_1, \cdots, p'_1}_{N'_1},
		\cdots,
		\underbrace{p'_{n_\text{out}}, \cdots, p'_{n_\text{out}}}_{N'_{n_\text{out}}}
		}^{N'}
	\bigg).
\end{split}
\end{align}
Using this, we can connect all loose ends by writing
\begin{align}
\begin{split}
&	\left[\prod_{r=1}^{N+N'}\frac{1}{2}\int \td{k_r}S^{a_r}_{\mu\nu}I^{\mu\nu\rho_r\sigma_r}\right]
	M^{(N+N')}_{\rho_1\sigma_1\cdots\rho_{N+N'}\sigma_{N+N'}}(k_1,\cdots,k_{N+N'}),
\end{split}
\end{align}
where $S^{a_r}_{\mu\nu}\equiv S_{\mu\nu}(a_r,k)$.
Writing out the expression for $M^{(N+N')}_{\rho_1\sigma_1\cdots}$, we obtain
\begin{align}
\label{eqn:comb_before}
\begin{split}
	&
	(-1)^N
	\sum_{s=0}^{N+N'}
	\left[\prod_{r=1}^{N+N'}\frac{1}{2}\int \td{k_r}S^{a_r}_{\mu\nu}I^{\mu\nu\rho_r\sigma_r}\right]
	\\ &\qquad\times
	\sum_{\substack{\text{Perm} \\ (\rho,\sigma,k)}}
	\left[\frac{1}{s!}\prod_{i=1}^sP^\text{tot}_{\rho_i\sigma_i}(k_i)\right]
	\frac{\xi_{\rho_{s+1}\sigma_{s+1}\cdots\rho_{N+N'}\sigma_{N+N'}}(k_{s+1},\cdots,k_{N+N'})}{(N+N'-s)!}.
\end{split}
\end{align}
The summand of $\sum_s$ is the sum of all diagrams where $s$ of the $N+N'$ interacting gravitons
	are being connected to external legs.
For a given $s$, let us say there are $s_i$ ($s'_i$) gravitons connecting the $i$th ($j$th) incoming (outgoing)
	cloud to external legs, so that $\sum_{i\in\text{in}}s_i+\sum_{j\in\text{out}}s'_j=s$.
Then, instead of summing over the total number $s$, we can sum over each of the numbers $s_i$ an $s'_i$.
This yields
\begin{align}
\begin{split}
	&
	(-1)^N
	\left[\prod_{i\in\text{in}}\sum_{s_i=0}^{N_i}\right]
	\left[\prod_{j\in\text{out}}\sum_{s'_j=0}^{N'_j}\right]
	\left[\prod_{r=1}^{N+N'}\frac{1}{2}\int \td{k_r}S^{a_r}_{\mu\nu}I^{\mu\nu\rho_r\sigma_r}\right]
	\\ &\qquad\times
	\sum_{\substack{\text{Perm} \\ (\rho,\sigma,k)}}
	\left[\frac{1}{s!}\prod_{i=1}^sP^\text{tot}_{\rho_i\sigma_i}(k_i)\right]
	\frac{\xi_{\rho_{s+1}\sigma_{s+1}\cdots\rho_{N+N'}\sigma_{N+N'}}(k_{s+1},\cdots,k_{N+N'})}{(N+N'-s)!},
\end{split}
\end{align}
where $s$ is now defined as $s=\sum_{i\in\text{in}}s_i+\sum_{j\in\text{out}}s'_j$.
Among the $N+N'$ copies of $\frac{1}{2}\int \td{k} S_{\mu\nu}I^{\mnrs}$, $s$ copies will contract with
	$\prod P_{\rho\sigma}$ (corresponding to external legs) and form the IR-divergent factor; the remaining
	$N+N'-s$ copies will contract with $\xi_{\rho\sigma\cdots}$ and end up in the IR-convergent part.
For the $i$th incoming cloud, there are $N_i$ copies of $\frac{1}{2}\int \td{k} S_{\mu\nu}I^{\mnrs}$
	but only $s_i$ copies of $\prod P^\text{tot}_{\rho\sigma}$, which indicates that we get ${N_i \choose s_i}$ identical
	contractions.
By the same token, the $j$th outgoing cloud has ${N'_j \choose s'_j}$ identical contractions.
Therefore, contracting the indices and distributing $(-1)^N$ yields
\begin{align}
\label{eqn:factered_out}
\begin{split}
	&
	\left[\prod_{i\in\text{in}}\sum_{s_i=0}^{N_i}{N_i \choose s_i}
	\left(-\frac{1}{2}\int \td{k}S^i_{\mu\nu}I^{\mnrs}P^\text{tot}_{\rho\sigma}\right)^{s_i}\right]
	\left[\prod_{j\in\text{out}}\sum_{s'_j=0}^{N'_j}{N'_j \choose s'_j}
	\left(\frac{1}{2}\int \td{k}S^j_{\mu\nu}I^{\mnrs}P^\text{tot}_{\rho\sigma}\right)^{s'_j}\right]
	\\ &\qquad\times
\Qfintilde_{N_1-s_1,\cdots,N_{n_\text{in}}-s_{n_\text{in}},N'_1-s'_1,\cdots,N'_{n_\text{out}}-s'_{n_\text{out}}}
\ ,
\end{split}
\end{align}
where $\Qfintilde_{N_1-s_1,\cdots}$ is the IR-convergent part of the amplitude (from the contractions with $\xi$),
	given by
\begin{align}
\begin{split}
\Qfintilde&_{N_1-s_1,\cdots,N_{n_\text{in}}-s_{n_\text{in}},
	N'_1-s'_1,\cdots,N'_{n_\text{out}}-s'_{n_\text{out}}}
= (-1)^{\sum_{i\in\text{in}}(N_i-s_i)}
\\	&\times
\left[\prod_{r=1}^{N+N'-s}\frac{1}{2}\int \td{k_r}S^{a'_r}_{\mu\nu}I^{\mu\nu\rho_r\sigma_r}\right]
\xi_{\rho_1\sigma_1\cdots\rho_{N+N'-s}\sigma_{N+N'-s}}(k_1,\cdots,k_{N+N'-s}).
\end{split}
\end{align}
Here we used a sequence $a'$ similar to \eqref{eqn:seq} to simplify the notation:
\begin{align}
\big(a'_i\big)_{i=1}^{N+N'-s}
	&\equiv\bigg(
		\overbrace{
		\underbrace{p_1, \cdots, p_1}_{N_1-s_1},
		\cdots,
		\underbrace{p_{n_\text{in}}, \cdots, p_{n_\text{in}}}_{N_{n_\text{in}}-s_{n_\text{in}}},
		}^{N-\sum_is_i}
		\overbrace{
		\underbrace{p'_1, \cdots, p'_1}_{N'_1-s'_1},
		\cdots,
		\underbrace{p'_{n_\text{out}}, \cdots, p'_{n_\text{out}}}_{N'_{n_\text{out}}-s'_{n_\text{out}}}
		}^{N'-\sum_js'_j}
	\bigg).
\end{align}
The first line of \eqref{eqn:factered_out} is the IR-divergent contribution of the configuration
	({$\lbrace N_i\rbrace$},$\lbrace N'_j\rbrace$) factored out of the amplitude.
	
There is a combinatorial factor that accompanies \eqref{eqn:factered_out},
	and to compute this we have to take into account other types of contributing gravitons.
A cloud has three types of gravitons attached to it: the interacting gravitons, the disconnected gravitons,
	and the in-to-in/out-to-out gravitons.
The in-to-in and out-to-out gravitons will later be treated separately,
	so for the moment let us assume that there are only the first two types.
Let $l_i$ ($l'_j$) denote the number of disconnected gravitons attached to a cloud of an incoming (outgoing) scalar.
$l$ is the total number of disconnected graviton lines, so that $\sum_i l_i=\sum_j l'_j = l$.
A cloud with $N_i$ interacting and $l_i$ disconnected
	graviton lines attached to it involves $N_i+l_i$ graviton creation/annihilation operators,
	which means it comes from the ($N_i+l_i$)-th term in the Taylor expansion of $e^{\pm R_f(p)}$.
	This term is accompanied by the factor $1/(N_i+l_i)!$.
Since there are ${N_i +l_i\choose N_i}=(N_i+l_i)!/l_i!N_i!$ ways to group these into interacting/disconnected gravitons,
	this cloud has a net factor of $1/l_i!N_i!$.
This applies to every incoming and outgoing cloud,
	and therefore the configuration ($\lbrace N_i, l_i\rbrace$, $\lbrace N'_j,l'_j\rbrace$) has a net combinatorial factor of
\begin{align}
\label{eqn:int_nonint_comb}
\left[\prod_{i\in\text{in}}\frac{1}{l_i!N_i!}\right]
\left[\prod_{j\in\text{out}}\frac{1}{l'_j!N'_j!}\right].
\end{align}	
Multiplying this with \eqref{eqn:factered_out} yields
\begin{align}
\begin{split}
\label{eqn:fac_and_int}
&\left[\prod_{i\in\text{in}}\frac{1}{l_i!}\sum_{s_i=0}^{N_i}\frac{1}{s_i!}
\left(-\frac{1}{2}\int \td{k}S^i_{\mu\nu}I^{\mnrs}P^\text{tot}_{\rho\sigma}\right)^{s_i}\right]
\left[\prod_{j\in\text{out}}\frac{1}{l'_j!}\sum_{s'_j=0}^{N'_j}\frac{1}{s'_j!}
\left(\frac{1}{2}\int \td{k}S^j_{\mu\nu}I^{\mnrs}P^\text{tot}_{\rho\sigma}\right)^{s'_j}\right]
\\	&\qquad\times	
	\Qfin_{N_1-s_1,\cdots,N_{n_\text{in}}-s_{n_\text{in}},N'_1-s'_1,\cdots,N'_{n_\text{out}}-s'_{n_\text{out}}},
\end{split}
\end{align}
where
\begin{align}
\Qfin_{j_1,j_2,\cdots,j_{n_\text{in}},j'_1,j'_2,\cdots,j'_{n_\text{out}}}
	\equiv \frac{\Qfintilde_{j_1,j_2,\cdots,j_{n_\text{in}},j'_1,j'_2,\cdots,j'_{n_\text{out}}}}
	{j_1!j_2!\cdots j_{n_\text{in}}! j'_1!j'_2!\cdots j'_{n_\text{out}}!}
\end{align}
is the the rescaled finite amplitude.

We also have the contribution from the disconnected gravitons.
A graviton line connecting the $i$th incoming cloud to the $j$th outgoing cloud contributes a factor
\begin{align}
\frac{1}{2}\int \td{k} S_{\mu\nu}(p'_j,k)I^{\mnrs}S_{\rho\sigma}(p_i,k).
\end{align}
Summing over all possible disconnected lines therefore contributes the factor
\begin{align}
\label{eqn:nonint}
l!\left[\frac{1}{2}
	\sum_{\substack{n\in\text{out}\\m\in\text{in}}}
	\int \td{k}S^n_{\mu\nu}I^{\mnrs}S^m_{\rho\sigma}\right]^l
=l!\left[\frac{1}{2}\int \td{k}
	S^\text{out}_{\mu\nu}I^{\mnrs}S^\text{in}_{\rho\sigma}\right]^l,
\end{align}
where $l!$ is the number of ways we can pair $l$ incoming gravitons with $l$ outgoing gravitons.

The product of \eqref{eqn:fac_and_int} and \eqref{eqn:nonint} form the contribution of a single configuration
	($\lbrace N_i,l_i\rbrace$, $\lbrace N'_j,l'_j\rbrace$).
Taking these two expressions and summing over all $N_i$, $N'_j$, $l_i$, $l'_j$, and $l$ gives us the amplitude
\begin{align}\label{eqn:b4rearrange}
\begin{split}
&\sum_{l=0}^{\infty}\sum_{\sum l_i=l}\sum_{\sum l'_j=l}\sum_{\lbrace N_i\rbrace}\sum_{\lbrace N'_j\rbrace}
l!\left[\frac{1}{2}\int \td{k}S^\text{out}_{\mu\nu}I^{\mnrs}S^\text{in}_{\rho\sigma}\right]^l
\\&\times	
\left[\prod_{i\in\text{in}}\frac{1}{l_i!}\sum_{s_i=0}^{N_i}\frac{1}{s_i!}
\left(-\frac{1}{2}\int \td{k}S^i_{\mu\nu}I^{\mnrs}P^\text{tot}_{\rho\sigma}\right)^{s_i}\right]
\\&\times	
\left[\prod_{j\in\text{out}}\frac{1}{l'_j!}\sum_{s'_j=0}^{N'_j}\frac{1}{s'_j!}
\left(\frac{1}{2}\int \td{k}S^j_{\mu\nu}I^{\mnrs}P^\text{tot}_{\rho\sigma}\right)^{s'_j}\right]
\Qfin_{N_1-s_1,\cdots,N_{n_\text{in}}-s_{n_\text{in}},N'_1-s'_1,\cdots,N'_{n_\text{out}}-s'_{n_\text{out}}}
.
\end{split}
\end{align}
Let us use the identity
\begin{align}
\sum_{\sum l_i=l}\frac{l!}{l_1!l_2!\cdots l_{n_\text{in}}!}=1
\end{align}
to eliminate the $l_i$'s and $l'_j$'s, and rearrange the sums
\begin{align}
\sum_{N_i=0}^\infty\sum_{s_i=0}^{N_i}
\quad\rightarrow\quad \sum_{s_i=0}^\infty\sum_{N_i=s_i}^{\infty}
\quad\rightarrow\quad \sum_{s_i=0}^\infty\sum_{m_i=0}^{\infty}
\end{align}
with $m_i\equiv N_i-s_i$, after which \eqref{eqn:b4rearrange} becomes
\begin{align}
\begin{split}
&\sum_{l=0}^{\infty}
\frac{1}{l!}\left[\frac{1}{2}\int \td{k}S^\text{out}_{\mu\nu}I^{\mnrs}S^\text{in}_{\rho\sigma}\right]^l
\left[\prod_{i\in\text{in}}\sum_{s_i=0}^{\infty}\sum_{m_i=0}^\infty\frac{1}{s_i!}
\left(-\frac{1}{2}\int \td{k}S^i_{\mu\nu}I^{\mnrs}P^\text{tot}_{\rho\sigma}\right)^{s_i}\right]
\\	&\qquad\times	
\left[\prod_{j\in\text{out}}\sum_{s'_j=0}^{\infty}\sum_{m'_j=0}^\infty\frac{1}{s'_j!}
\left(\frac{1}{2}\int \td{k}S^j_{\mu\nu}I^{\mnrs}P^\text{tot}_{\rho\sigma}\right)^{s'_j}\right]
\Qfin_{m_1,\cdots,m_{n_\text{in}},m'_1,\cdots,m'_{n_\text{out}}}
.
\end{split}
\end{align}
The divergent factors exponentiate, leaving us with
\begin{align}
\label{eqn:after_exp}
\begin{split}
&\exp\left(
\frac{1}{2}\int \td{k}S^\text{out}_{\mu\nu}I^{\mnrs}S^\text{in}_{\rho\sigma}
+\frac{1}{2}\int \td{k}S^\text{tot}_{\mu\nu}I^{\mnrs}P^\text{tot}_{\rho\sigma}\right)
\Qfin\ ,
\end{split}
\end{align}
where the leftover, IR-finite part $\Qfin$ of the amplitude is given by
\begin{align}
\Qfin\equiv \left[\prod_{i\in\text{in}}\sum_{m_i=0}^\infty\right]\left[\prod_{j\in\text{out}}\sum_{m'_j=0}^\infty\right]
\Qfin_{m_1,\cdots,m_{n_\text{in}},m'_1,\cdots,m'_{n_\text{out}}}\ .
\end{align}

Now we consider the contribution of the in-to-in and out-to-out gravitons.
These contributions manifest themselves in the form of normalization of the in- and out-states.
In the single-particle case, we used the BCH formula to discard the annihilation operators.
We should be more careful in doing so when dealing with the general case of multi-particle state.
Consider for example the following two-particle state:
\begin{align}
\ket{\text{i}} &=e^{R_f(p_1)}b^\dagger(p_1)e^{R_f(p_2)}b^\dagger(p_2) \ket{0}\\
\begin{split}
\label{eqn:2ptl_in_state}
&=\exp\left(
		\int\td{k}
		S^1_{\mu\nu}a^{\dagger\mu\nu}
	\right)
	\exp\left(
		-\int\td{k}
		S^1_{\mu\nu}a^{\mu\nu}
	\right)
	\exp\left(
		-\frac{1}{4}\int \td{k}
		S^1_{\mu\nu}I^{\mnrs}S^1_{\rho\sigma}
	\right)
	\\&\times
	\exp\left(
		\int\td{k}
		S^2_{\mu\nu}a^{\dagger\mu\nu}
	\right)
	\exp\left(
		-\frac{1}{4}\int \td{k}
		S^2_{\mu\nu}I^{\mnrs}S_{\rho\sigma}^2
	\right)
	b^\dagger(p_1)b^\dagger(p_2)\ket{0}.
\end{split}
\end{align}
We wish to eliminate $\exp\left(-\int\td{k}S^1_{\mu\nu}a^{\mu\nu}\right)$ in \eqref{eqn:2ptl_in_state}, by commuting it all the way to the vacuum;
but it does not commute with $\exp\left(\int\td{k}S^2_{\mu\nu}a^{\dagger\mu\nu}\right)$, and thus this procedure induces an extra factor.
Since
\begin{align}
e^Ae^B=e^{A+B}e^{\frac{1}{2}[A,B]}=e^{B+A}e^{\frac{1}{2}[A,B]}=e^Be^Ae^{[A,B]}
\end{align}
for $[A,B]\in\mathbb{C}$, the extra factor is
\begin{align}
&\exp\left\lbrace
	-\int\td{k}\td{k'}
	S_{\mu\nu}(p_1,k)S_{\rho\sigma}(p_2,k')\left[a^{\mu\nu}(k),a^{\dagger\rho\sigma}(k')\right]
\right\rbrace\\
&=\exp\left\lbrace
	-\frac{1}{2}\int \td{k}S_{\mu\nu}(p_1,k)I^{\mnrs}S_{\rho\sigma}(p_2,k)
\right\rbrace\\
&=\exp\left\lbrace
	-\frac{1}{4}\int \td{k}
	\left(
		S^1_{\mu\nu}I^{\mnrs}S^2_{\rho\sigma}
		+S^2_{\mu\nu}I^{\mnrs}S^1_{\rho\sigma}
	\right)
\right\rbrace,
\end{align}
where in the last line we used the symmetry of $I^{\mnrs}$ to write the expression in a symmetric fashion.
For a multi-particle in-state, we will get a factor of this form for each
unordered pair of incoming scalars.
With a similar line of reasoning for a multi-particle out-state, the total contribution
of the in-to-in and out-to-out gravitons will result in a factor of
\begin{align}
\label{eqn:normfact}
\exp\left\lbrace
-\frac{1}{4}\int \td{k}\left(
S^\text{in}_{\mu\nu}I^{\mnrs}S^\text{in}_{\rho\sigma}
+S^\text{out}_{\mu\nu}I^{\mnrs}S^\text{out}_{\rho\sigma}
\right)
\right\rbrace .
\end{align}
Multiplying \eqref{eqn:normfact} with the divergent factor in \eqref{eqn:after_exp} gives us
the expression for the net divergent factor $\Afactor_\text{cloud}$ due to the amplitude interactions involving the clouds.
\begin{align}
\begin{split}
\Afactor_\text{cloud}
\label{eqn:acloud_factor}
&=\exp\Bigg\lbrace
-\frac{1}{4}\int \td{k}\left(
S^\text{in}_{\mu\nu}I^{\mnrs}S^\text{in}_{\rho\sigma}
+S^\text{out}_{\mu\nu}I^{\mnrs}S^\text{out}_{\rho\sigma}\right)
\\&\qquad
+\frac{1}{2}\int \td{k}S^\text{out}_{\mu\nu}I^{\mnrs}S^\text{in}_{\rho\sigma}
+\frac{1}{2}\int \td{k} S^\text{tot}_{\mu\nu}I^{\mnrs}
P_{\rho\sigma}^\text{tot}
\Bigg\rbrace
\end{split}\\
&=\exp\left(
-\frac{1}{4}\int \td{k} S^\text{tot}_{\mu\nu}I^{\mnrs}S^\text{tot}_{\rho\sigma}
+\frac{1}{2}\int \td{k} S^\text{tot}_{\mu\nu}I^{\mnrs}P_{\rho\sigma}^\text{tot}
\right).
\end{align}
Since $S^\text{tot}_{\mu\nu}(k)=P^\text{tot}_{\mu\nu}(k)+C^\text{tot}_{\mu\nu}(k)$,
\begin{align}
\begin{split}
\Afactor_\text{cloud}
&=\exp\bigg\lbrace
\frac{1}{4}\int \td{k}\Big[
-\left(P^\text{tot}_{\mu\nu}+C^\text{tot}_{\mu\nu}\right)
I^\mnrs
\left(P^\text{tot}_{\rho\sigma}+C^\text{tot}_{\rho\sigma}\right)
+2\left(P^\text{tot}_{\mu\nu}+C^\text{tot}_{\mu\nu}\right)
I^\mnrs
P_{\rho\sigma}^\text{tot}
\Big]
\bigg\rbrace
\end{split}
\\
&=\exp\left\lbrace\frac{1}{4}\int \td{k}\left(
P^\text{tot}_{\mu\nu}I^{\mnrs}P^\text{tot}_{\rho\sigma}
-C^\text{tot}_{\mu\nu}I^{\mnrs}C^\text{tot}_{\rho\sigma}
\right)
\right\rbrace
\\
\label{eqn:cloudfinexp}
&=\exp\left(
\frac{1}{4}\sum_{n,m}\eta_n\eta_m\int \td{k}P^m_{\mu\nu}I^{\mnrs}P^n_{\rho\sigma}
\right)
\exp\left(
-\frac{1}{4}\int \td{k}
C^\text{tot}_{\mu\nu}I^{\mnrs}C^\text{tot}_{\rho\sigma}
\right)
\end{align}
The first exponential of \eqref{eqn:cloudfinexp} is the inverse of $\Afactor_\text{virt}$, so let us write
\begin{align}
\label{Acloudvirt}
\Afactor_\text{cloud}&=(\Afactor_\text{virt})^{-1} \exp(-a C)
\end{align}
where $a=\kappa^2/256\pi^2$ and
\begin{align}
\label{eqn:final_Ctot}
C\equiv \int \frac{d^3k}{\wk^3}
\left(
	\sum_{j\in\text{out}}c^j_\mn - \sum_{i\in\text{in}}c^i_\mn
\right)
I^\mnrs
\left(
	\sum_{j\in\text{out}}c^j_\rs - \sum_{i\in\text{in}}c^i_\rs
\right).
\end{align}
The factor $e^{-aC}$ derives solely from the interactions between graviton clouds.
It only contributes to the normalization of states, and we can use \eqref{eqn:cc3} to set $C=0$.
Therefore, $\Afactor_\text{cloud}$ exactly cancels the divergent
	factor $\Afactor_\text{virt}$, proving the cancellation of IR divergence to all orders.

Lastly, let us consider the general case where we use different dressings for the incoming and outgoing state.
Then, the expression \eqref{eqn:final_Ctot} readily generalizes to
\begin{align}
C\equiv \int \frac{d^3k}{\wk^3}
c^\text{tot}_\mn(k)
I^\mnrs
c^\text{tot}_\rs(k),
\quad\text{with}\quad
c_\mn^\text{tot}(k)
	\equiv
	\sum_{j\in\text{out}}c'_\mn(p_j,k)
	- \sum_{i\in\text{in}}c_\mn(p_i,k).
\end{align}
If $c^\text{tot}_\mn(k)$ does not vanish as $k\rightarrow 0$, then $C$ diverges and $e^{-aC}\rightarrow 0$,
	forcing the amplitude to be zero.
Non-zero amplitudes are therefore only allowed between asymptotic states whose c-matrices satisfy
\begin{align}
\label{eqn:ctot_must_die}
	\sum_{j\in\text{out}}c'_\mn(p_j,k)
	= \sum_{i\in\text{in}}c_\mn(p_i,k),
\end{align}
up to subleading corrections of order $O(k)$.

\section{BMS modes}\label{GaugeModesApp}

In this appendix we review and extend on the work of \cite{Campiglia:2015kxa, Campiglia:2015lxa, Campiglia:2016efb}
	on the solutions of the pure gauge mode $\lambda_{\mu}$.
The Laplace equation \eqref{LaplaceEq} in the hyperbolic system of coordinates takes the form
\begin{equation}
\pa_{\nu}\pa^{\nu} \lambda_{\mu} = \left(  \frac{\triangle_{\rho}}{\tau^2}  -  \pa_{\tau}^2  \right) \lambda _{\mu} =0
\ ,
\end{equation}
where
\begin{equation}
\triangle_{\rho} = (1+\rho ^2 ) \pa_{\rho}^2 + \frac{1}{\rho} (2+3\rho^2) \pa_{\rho} + \frac{1}{\rho ^2} (1+z \zb)^2 \pa_{z} \pa_{\zb}
\ .
\end{equation}
At $\tau \rightarrow \infty$ the only non-vanishing component of $\lambda_{\mu}$
	is $\lambda_{\tau}$,
\begin{equation}
\lim_{\tau \rightarrow \infty }\lambda_{\tau}(\tau,\rho,z,\zb) = \lamt_{\tau} (\rho,z,\zb)
\ ,
\end{equation}
the asymptotic form of which obeys the following equation
\begin{equation}
\triangle_{\rho}  \lamt _{\tau} = n (n-2)  \lamt _{\tau} 
\ ,
\end{equation}
where $n=3$ in our case (for a $U(1)$ gauge symmetry, $n=2$).
$\lamt_{\tau} (\rho,z,\zb)$ can be written in terms of the Green's function
\begin{equation}
\lamt_{\tau} (\rho,z,\zb)  = \int   d^2 \omega \, G(\rho,z,\zb ; \omega , \omegab) f(\omega,\omegab)
\ .
\end{equation}
The Green's function obeys
\begin{eqnarray}
\label{GreenEq}
 &\triangle_{\rho}  G = n (n-2)  G  \\
 \label{GreenNormal}
& \lim_{\rho \rightarrow \infty} 
 \rho ^{ 2-n} G  (\rho,z,\zb ; \omega , \omegab) 
 =
 \delta^2 (z-\omega)
\ .
\end{eqnarray}
The two solutions to equation \eqref{GreenEq} are given by
\begin{equation}
G(\rho,z,\zb ; \omega , \omegab) = \alpha f^{(n)}(\rho,z,\zb ; \omega , \omegab)  + \beta f^{(2-n)}(\rho,z,\zb ; \omega , \omegab) 
\end{equation}
where
\begin{equation}
 f^{(n)}(\rho,z,\zb ; \omega , \omegab)  = \frac{n-1}{2^{n-1}} \frac{\sqrt{\gamma}}{2\pi} \left(  \sqrt{1+\rho^2} - \rho \, \hat{x}_z \cdot \hat{x}_{\omega} \right)^{-n}.
\end{equation}
The asymptotic of the function $ f^{(n)}$ is
\begin{equation}
\lim_{\rho \rightarrow \infty } f^{(n)}(\rho,z,\zb ; \omega , \omegab)
\sim
\begin{cases}
\rho^{-n} ,\qquad \hat{x}_z \neq \hat{x}_{\omega}  \\
\rho^{+n} ,\qquad \hat{x}_z = \hat{x}_{\omega} 
\end{cases}
\end{equation}
and its integral over $S^2$ asymptotes to
\begin{equation}
\lim_{\rho \rightarrow \infty }  \rho^{2-n} \int d^2 \omega \, f^{(n)}(\rho,z,\zb ; \omega , \omegab) =1
\ .
\end{equation}
The solution for the gauge mode therefore asymptotes to
\begin{equation}\label{lambdaSol}
\begin{aligned}
\lim_{\rho \rightarrow \infty } \lamt_{\tau} (\rho,z,\zb) 
& =
 \alpha (z,\zb) \, \rho^{n-2}   \left( 1 + \dots  \right) 
+\beta (z,\zb) \, \rho^{-n} \left( 1 + \dots   \right) \ ,
\end{aligned}
\end{equation}
where the dots stand for subleading terms in $1/\rho$.
The $\alpha$-series is leading and do not vanish at time-like infinity $\rho \rightarrow \infty$. It is a Large Gauge Transformation. The $\beta$-series is subleading and vanishes at time-like infinity.

We would now like to express the $\alpha$ and $\beta$ modes in terms of the radiative data $C_{zz}$ and $C_{\zb\zb}$. To do this we should study the solutions to equation \eqref{xiEq} for the gauge mode $\xi _{\mu}$. At leading order, only the $\tau$-component is non-vanishing and its solution can be written in terms of the Green's function
\begin{equation}
\xi _{\tau}
=
\int d^2 \omega \, G(\rho,z,\zb; \omega , \omegab)
\left(
\frac{1}{2} \pa_{\tau} h^B - \pa^{\nu} h^B_{\tau \nu}
\right)
\ .
\end{equation}
Plugging the solution for the Green's function, and the asymptotic form of the metric in the Bondi gauge, we get for the $\alpha$-mode
\begin{equation}
\lim_{\rho \rightarrow \infty}\xi _{\tau}^{\alpha}
=
\rho^{n-2}
\left(
\pa_z U_{\zb} + \pa_{\zb} U_z
\right)_{\mI^+_+}
\ .
\end{equation}
By comparing to \eqref{lambdaSol}, we conclude that
\begin{equation}\label{amode}
\alpha =  \left( \pa_z U_{\zb} + \pa_{\zb} U_z \right)_{\mI^+_+}
\ .
\end{equation}
To solve for the subleading $\beta$-mode in a similar way we have to study subleading corrections to $\xi^{\alpha}$.
Here, we will not solve this problem explicitly, but instead give a heuristic explanation based on properties of 2D conformal field theories. On $S^2$ the leading $\alpha$-mode is a left mover, while the subleading $\beta$-mode is a right mover.
This implies that the $\beta$-mode is orthogonal to \eqref{amode} and is therefore given by
\begin{equation}
\beta = i \left( \pa_z U_{\zb} - \pa_{\zb} U_z \right)_{\mI^+_+}
\ .
\end{equation}
The factor of $i$ is required to make $\beta$ real.
We leave the explicit analysis and further exploration of this direction to future work.
See \cite{Cheung:2016iub} for a related work.

\bibliographystyle{ssg}


\bibliography{SoftGravitons.bib}

\end{document}